\documentclass{ws-jtca}

\usepackage{amsmath}
\usepackage{multirow}
\usepackage{graphicx}
\usepackage[caption=false]{subfig}
\usepackage{placeins}
\usepackage{booktabs,tabularx}
\usepackage{bbold}
\usepackage{rotating}

\usepackage{listings}
\usepackage{color}

\usepackage{lineno}
\definecolor{dkgreen}{rgb}{0,0.6,0}
\definecolor{gray}{rgb}{0.5,0.5,0.5}
\definecolor{mauve}{rgb}{0.58,0,0.82}

\newcommand{\eq}[1]{\begin{equation}#1\end{equation}}
\newcommand{\al}[1]{\begin{equation}\begin{aligned}#1\end{aligned}\end{equation}}
\newcommand{\nn}{\nonumber}
\def\pd{\partial}
\newcommand{\pdif}[2]{\frac{\pd #1}{\pd #2}}

\newcolumntype{Y}{>{\centering\arraybackslash}X}

\newcommand{\bea}{\begin{eqnarray}}
\newcommand{\eea}{\end{eqnarray}}

\AtBeginDocument{%
  \heavyrulewidth=.08em
  \lightrulewidth=.05em
  \cmidrulewidth=.03em
  \belowrulesep=.65ex
  \belowbottomsep=0pt
  \aboverulesep=.4ex
  \abovetopsep=0pt
  \cmidrulesep=\doublerulesep
  \cmidrulekern=.5em
  \defaultaddspace=.5em
}

\begin{document}

%
\catchline{}{}{}{}{}
%

\title{Stable rational approximations for parabolic equation methods}

\author{Adith Ramamurti}
\address{Code 7160, Acoustics Division, \\U.S. Naval Research Laboratory, \\Washington, DC 20375 \\ adith.ramamurti.civ@us.navy.mil}
\author{Joseph F. Lingevitch}
\address{Code 7160, Acoustics Division, \\U.S. Naval Research Laboratory, \\Washington, DC 20375}
\author{Jonathan C. Lighthall}
\address{Code 7180, Acoustics Division, \\U.S. Naval Research Laboratory, \\Stennis Space Center, Mississippi 39529}
\author{Michael D. Collins}
\address{Code 7160, Acoustics Division, \\U.S. Naval Research Laboratory, \\Washington, DC 20375}

\maketitle


\newpage


\begin{abstract}
Modern parabolic equation (PE) methods for wave propagation rely on application of a variety of fractional-powered differential operators. 
Rational approximations of these operators need to properly map their spectra onto the complex plane, 
accurately handling propagating modes while annihilating evanescent ones.
Standard approaches for stable and accurate rational approximations include rotating the branch cut of the operators or imposing stability constraint equations,
and have yielded accurate results for wave propagation in a variety of fluid, elastic, and fluid-elastic waveguides.
The stability constraint method, however, does not yield operators that are stable for all fluid-elastic waveguides, and
a recent study of waveguides comprised of a thin elastic layer
overlaying a thick fluid layer revealed instabilities in the approximations derived from rotated operators.
In this paper, we demonstrate the applicability of a different rational approximation method, the recently-developed adaptive Antoulas-Anderson (AAA) algorithm,
to simulations of wave propagation using the fluid-elastic parabolic equation.
We find that simulations using operators approximated using the AAA algorithm provide excellent agreement with reference solutions, 
with errors in transmission loss comparable to, and often less than, that of simulations using the rotated operator method.
In addition, we find that the AAA algorithm allows for the application of the split-step Pad\'e method to fluid-elastic waveguides,
which yields a large gain in computational efficiency.
\end{abstract}

\keywords{Seismoacoustic propagation; parabolic equations; rational approximations.}

\newpage

\section{Introduction}

Parabolic wave equations (one-way wave equations) are an approximate method used to obtain accurate and efficient solutions 
for many problems in which a solution of the full wave equation is impractical\cite{leontovich1946,tappert1977}.
One common application of parabolic equation (PE) methods is to wave propagation in laterally-varying fluid and elastic media. 
In a two-dimensional layered medium parameterized by range ($r$) and depth ($z$),
the parabolic (one-way) wave equation takes the form
\eq{
\pdif{u}{r} = ik_0\left(-1+\sqrt{1+q}\right)u\,. \label{eq:pe}
}
for the field $u = pe^{-ik_0r}$, where $p$ is a scalar (fluid) or vector (elastic) field, $k_0$ is the reference wavenumber,
and $q$ is a depth operator involving depth derivatives\cite{tappert1977,collins2019.1}.

One approach for numerical solutions of this equation involves approximating the square root operator with either a Taylor series or rational approximation in $q$,
discretizing the operator $q$, and solving using a Crank-Nicolson marching scheme.
Another common approach, known as the split-step Pad\'e method, involves approximating the exponential operator in the analytic solution,
\eq{
u(r+\Delta r) = \exp\left\{ik_0\Delta r\left(-1+\sqrt{1+q}\right)\right\} u(r)\,, \label{eq:epe}
}
with a rational approximation in $q$. The operator $q$ is then discretized on a depth grid, as in the Crank-Nicolson solution, and solved using a marching scheme\cite{collins1993}.
The latter approach allows for significantly larger range steps, and therefore
significantly increased computational efficiency, as the $k_0\Delta r$ term is included in the coefficients of the rational approximation of the propagation operator.

The fields are typically initialized at a finite distance $r_0$ using a self-starter method\cite{collins1992}, which involves application of operators such as
\eq{
(1-iq)^2(1+q)^{-1/4}\exp\left\{ik_0r_0\left(-1+\sqrt{1+q}\right)\right\}\,.
}
Accurate handling of variation of the environment in the transverse direction $r$
is done by either enforcing energy conservation\cite{collins1993.1}, the formulation of which utilizes operators of the form
\eq{
(1+q)^{\pm1/4}\,,
}
or by a single-scattering calculation\cite{lingevitch2010,collins2012}, which uses the square root operator.

Approximations of these operators, however, are not straightforward.
Early attempts at solving the PE numerically approximated the square root with a Taylor expansion\cite{tappert1977}.
This, however, yields unstable behavior as the evanescent spectrum ($q<-1$) cannot be properly 
represented by a real-valued polynomial.
An improved approach involves using rational approximations of the square root\cite{bamberger1988,collins1989,wetton1990,collins1991.0,collins1991.1}, 
which, when slightly modified using stability constraint equations, annihilates the evanescent spectrum\cite{collins1991.0,collins1993}.
 This approach for stabilizing these operators, however, fails for certain cases of wave propagation in fluid and elastic waveguides\cite{wetton1990,collins1991.0,milinazzo1997}.

To attempt to rectify this issue, a new method using rotations of the branch cut of the square root operator was introduced
 by Milinazzo, et al.\cite{milinazzo1997}, and expanded upon by Lingevitch and Collins\cite{lingevitch1998} for the exponential operator.
The branch cut of the square root is rotated such that the eigenvalues of evanescent modes are forced into the upper half of the complex plane.
Since its introduction, this rotated operator Pad\'e (ROP) method has become a common method
for approximating the square root operator in PE simulations.

The ROP method works extremely well for the square root operator, but not as well for the exponential operator.
As a result, the exponential operator is typically stabilized using a constraint equation, which works excellently for the pure-fluid case,
but is unstable for some cases of propagation modeling in elastic and coupled fluid-elastic waveguides;
the significant boost in computational efficiency from the split-step Pad\'e method is therefore not available for all PE-based simulations.
 Nevertheless, PE simulations using the Crank-Nicolson approach are still significantly more 
 computationally efficient than computing full-field solutions to elastic and fluid-elastic wave equations.

In Appendices \ref{apx:ratapx}  and \ref{apx:hist}, we describe in detail these standard methods of calculating coefficients for and 
stabilizing the rational approximations of the generalized PE operator,
\eq{
f(q) = \exp\Big\{i\sigma\left(-1+\sqrt{1+q}\right) + \delta \ln(1+q) + \nu \ln(1+cq)\Big\}\,, \label{eq:peop}
} 
\eq{
f(q) \approx g(q) \equiv 1 + \sum_{j=1}^N \frac{\alpha_{j} q}{1+\mu_{j} q } = \prod_{j=1}^N \frac{1+\gamma_{j} q}{1+\mu_{j} q} \label{eq:pade}\,.
}

Recent work has extended the parabolic equation method to waveguides where a thin elastic layer
overlays a fluid\cite{collins2015,fialkowski2018}. 
As initially pointed out by Wetton and Brooke\cite{wetton1990}, the eigenvalues of the field propagating in thin elastic layers 
sometimes fall far below the real line on the complex plane, and can cause instabilities if not properly mapped to the upper half-plane. 

To stabilize these simulations (i.e. map these eigenvalues to the upper half-plane), 
the rotation angle of the branch cut of the square root operator must be increased to, and sometimes past, $90^\circ$. 
While this works well at low frequencies,
large rotations have the side effect of mapping the wavenumbers of some propagating modes in waveguides with thick fluid layers
into the lower half-plane, causing instability at higher frequencies; these propagating modes, which are high-angle, grow exponentially\cite{collins2019.2}.
This observation means that propagation modeling of these kinds of waveguides
may be unstable (or at best inaccurate) at higher frequencies when using the ROP method to approximate the PE operators.

In the decades since the previously discussed approaches were first applied to parabolic equation based wave propagation simulations 
and adopted as the standard methods, there have been many developments in the field of rational approximations. 
Among many others, these include vector fitting\cite{gustavsen1999,gustavsen2006,hendrickx2006} and RKFIT\cite{berljafa2015,berljafa2017} 
(see Sec. 11 of Nakatsukasa, et al.\cite{nakatsukasa2018} for additional discussion and references).
A significant advance in this area was the development of the adaptive Antoulas-Anderson (AAA) algorithm\cite{nakatsukasa2018}, which has had significant impact 
in a variety of fields\cite{nakatsukasa2023}.

In this paper, we demonstrate the applicability of the AAA algorithm for rational approximation to the operators present in parabolic equation methods,
with particular focus on propagation in waveguides where thin elastic layers overlay thick fluid layers.
In addition, we systematically study the effect of rotation angle of the ROP method on stability and accuracy of propagation modeling at higher frequencies.

\section{The AAA algorithm applied to PE operators}

The AAA algorithm yields rational approximations in barycentric form,
\eq{
f(q) \approx r^{(N+1)}(q) = \frac{n^{(N+1)}(q)}{d^{(N+1)}(q)} = \left(\sum_{j=1}^{N+1} \frac{w_j f_j}{q-q_j}\right)\Big/\left(\sum_{j=1}^{N+1} \frac{w_j}{q-q_j}\right)\,,\label{eq:bary}
}
where $\{q_j\}$ are support points, $\{f_j\}$ are data values, and $\{w_j\}$ are weights.
The algorithm for determining these values is detailed in Nakatsukasa, et al.\cite{nakatsukasa2018}; implementations
are available in as part of the Chebfun package\cite{chebfun}.

We now give a brief description of the AAA algorithm, which is an iterative procedure with $N+1$ steps.
We start with a finite set of points $Q\subset \mathbb{C}$.
At each step of the iteration $m$, the function $f(q)$ is represented by a rational approximation $r^{(m)}(q)$ of order $(m-1,~m-1)$ 
which interpolates the values in the set $\{f_j\}_m \equiv \{f(q_1),...,f(q_m)\}$, the function evaluated at the support points $\{q_j\}_m$.
We define the set $Q^{(m)}$ as the set of points $Q \char`\\ \{q_1,...,q_{m}\}$, the initial set $Q$ without the support points selected from steps 1 through $m$.

At step $m$, we select the next support point $q_m$,
where $q_m$ is the point in $Q^{(m-1)}$ that maximizes the residual $f(q)-n(q)/d(q)$, where $n(q)$ and $d(q)$ are of the form given in Eq. (\ref{eq:bary}).
The weights $w_1,...,w_m$ are then recomputed as to minimize $||f(q)d(q)-n(q)||_{Q^{(m)}}$ (the 2-norm over the set of points $Q^{(m)}$) with $||w||_m = 1$ (the 2-norm of the $m$-vector).

For implementation in the PE algorithms, the numerator and denominator can be expressed as products,
\eq{
\sum_{j=1}^{N+1} \frac{\alpha_j}{q+\beta_j} = \frac{\delta \prod_{j=1}^{N}(q+\eta_j)}{\prod_{j=1}^{N+1}(q+\beta_j)}\,,\nn
}
where $\delta = \sum \alpha_j$, $\alpha_j = w_j f_j$ for the numerator term,  $\alpha_j = w_j$ for the denominator term, and $\beta_j = -q_j$.

The coefficients $\{\eta\}$ can be determined by finding the $N$ roots of the polynomial
\eq{
\sum_{j=1}^{N+1} \alpha_j \prod_{i\neq j}^{N+1}(q+\beta_i)\,.\nn
} 

In product form, the denominators of the numerator and denominator terms of the barycentric form cancel, 
and therefore, the generalized PE operator can be expressed in a familiar form,
\eq{
f(q) \approx g(q) \equiv c \prod_{j=1}^N \frac{1+\gamma_{j} q}{1+\mu_{j} q} = c\left(1 + \sum_{j=1}^N \frac{\alpha_{j} q}{1+\mu_{j} q } \right) ,
}
with $\gamma_j = -1/\eta_{\text{num},j}$, $\mu_j = -1/\eta_{\text{den},j}$, and 
\eq{
c = \frac{\delta_\text{num}}{\delta_\text{den}} \prod_{j=1}^N \frac{\eta_{\text{num},j}}{\eta_{\text{den},j}}.\nn
}
For all forms of the PE operator given in Eq. (\ref{eq:peop}), $f(0)=1\rightarrow c=1$.
As this approximation is in the same form as the ``traditional'' Pad\'e approximant approach,
the computational cost of simulations using these approximations remains the same.

The strength of the AAA algorithm as applied to the PE operator is that the inputs to the algorithm are numerical,
rather than necessitating a $2N$-differentiable function as for computation of Pad\'e coefficients. 
As such, one can {\em explicitly} choose the mapping of inputs in the instances where the function has multiple branches. 
In the case of the square root operator, the lower half-plane can be chosen to map to the upper complex plane; 
for the exponential PE operator, one can map any values of $q$ with non-zero imaginary component to within the unit circle. 

In what follows, the set of input points $Q$ consists of 1200 evenly-spaced points on the real line from $-10$ to $10$.
For an isovelocity waveguide, the upper bound on real wave numbers is $k = k_r = \omega/c$, with $\omega$ the angular frequency
and $c$ the speed of sound (compressional or shear) in the medium\cite[p. 346]{jensen2011}.
The corresponding $Q_\text{max}$ is given by $c_0^2/c^2 - 1$, with $c_0$ the reference wave speed. 
Taking $Q_\text{max}=10$ covers the propagating wave number spectra of the operators for the examples presented in this work.
For waveguides having low shear-speed layers, for example, the upper limit of the interval would need to be increased to 
properly capture the propagating shear-wave spectrum.

The lower bound and sampling density of the interval was chosen after extensive testing;
the choice of lower endpoint covers the range of wave numbers in the evanescent spectrum
that contribute to the pressure and displacement fields in a variety of waveguides and frequencies. 
Errors were minimized when choosing a sampling density of 60 points per unit of the real line, regardless of choice of endpoints.

\begin{figure}
\centering
\subfloat[]{\includegraphics[width= .5\textwidth]{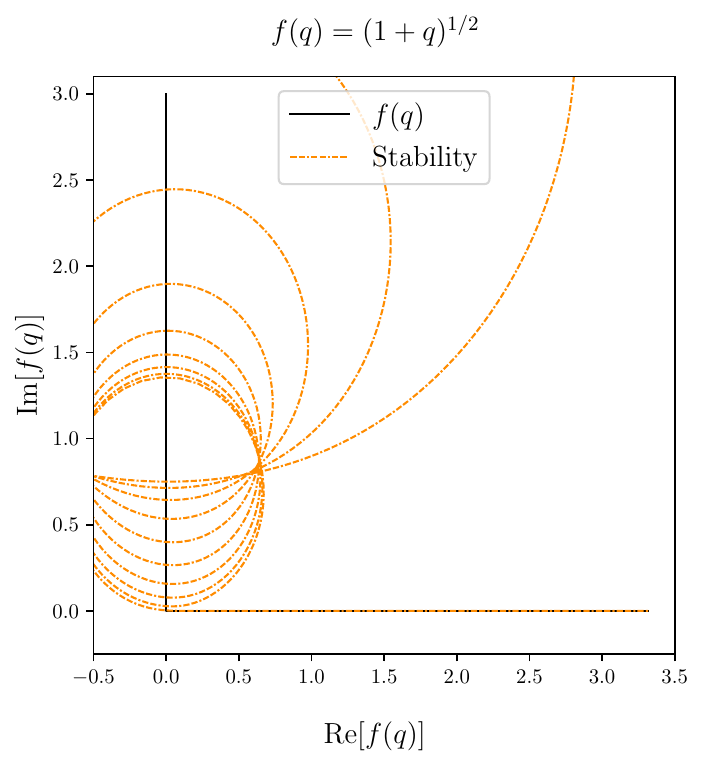}}
\subfloat[]{\includegraphics[width= .5\textwidth]{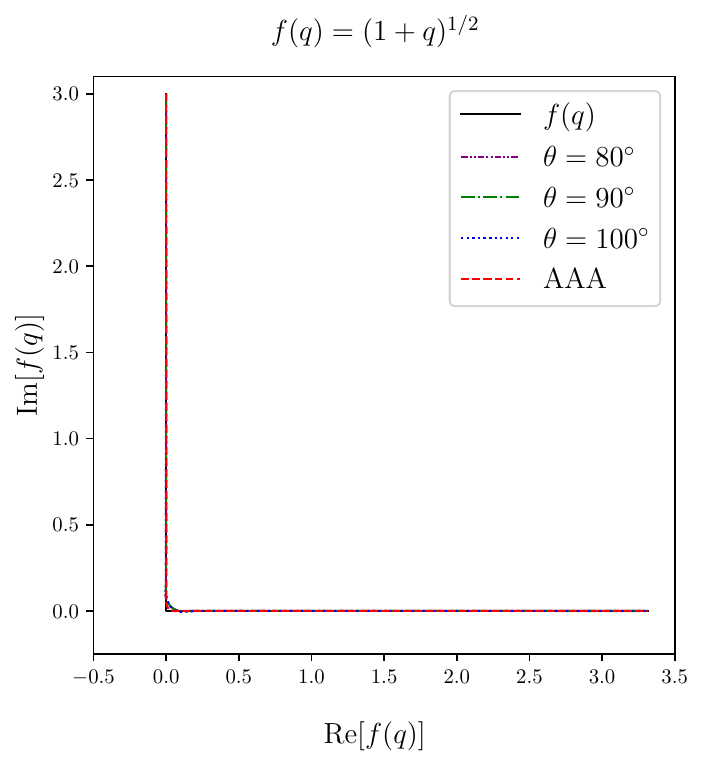}}
\caption{The mapping of the real line segment $(-10,10)$ onto the complex plane by the square root operator $\sqrt{1+q}$.
Shown (on both panels) are the the analytic result (solid black), and approximations using 
(a) the stability constraint method with a perturbation off the real line(dash-dash-dotted orange) 
and (b) the AAA algorithm (dashed red), 
the ROP method with $\theta=80^\circ$ (dashed-dot-dotted purple), $\theta=90^\circ$ (dash-dotted green),  $\theta=100^\circ$ (dotted blue).
All approximated operators are of order $N=12$.}
\label{fig:sqrt_comp}
\end{figure}

\begin{sidewaystable}
\centering
\vspace*{15cm} 
\caption{Product and sum coefficients for the square root operator for $N=12$ from the AAA algorithm.
These coefficients were determined using 1200 points evenly spaced on the line segment (-10,10).}
\vspace{1cm}
\resizebox{\textwidth}{!}{%
\begin{tabular}{c l l l}
\toprule
$j$ & $\alpha_j$ & $\gamma_j$ & $\mu_j$ \\ 
\midrule
1 & (6.3454575408843444e-02, -6.2042790452601387e-02) & (3.8177432536503741e-04, -1.9252093085902894e-02) & (5.9437515755782754e-05, -5.0082515094392755e-03) \\ 
2 & (6.0638207474204134e-02, -5.3563564229396010e-02) & (3.6754746991887141e-03, -7.1062397871649780e-02) & (1.3504655964612733e-03, -4.1376893133290536e-02) \\ 
3 & (6.9976957571447035e-02, -4.8363417013990107e-02) & (1.9575502449826566e-02, -1.5958578106573293e-01) & (8.8010069125327856e-03, -1.0960990838154383e-01) \\ 
4 & (9.6609535074277900e-02, -3.7662442346608607e-02) & (8.2984589300328246e-02, -3.0455565703321941e-01) & (4.1261276639561417e-02, -2.2399675581132486e-01) \\ 
5 & (1.2798452358218801e-01, 5.9040701414459018e-03) & (2.8289831019167805e-01, -4.8607580379400012e-01) & (1.5862404769444455e-01, -3.9748226124470942e-01) \\ 
6 & (9.5387370303509925e-02, 8.3217793291819095e-02) & (6.4649015137618682e-01, -5.2059407563834081e-01) & (4.5612610724523611e-01, -5.3722997756442248e-01) \\ 
7 & (7.9326601735299779e-03, 7.5615548905739363e-02) & (9.0973682476367523e-01, -3.3806169224543331e-01) & (8.0567620803191353e-01, -4.4197338531662417e-01) \\ 
8 & (-1.2601668081390334e-02, 2.7387856025578715e-02) & (9.9110742518513695e-01, -1.6389684097404239e-01) & (9.6574342448563255e-01, -2.4085596583576657e-01) \\ 
9 & (-6.5022520942920698e-03, 7.1531384935196434e-03) & (1.0028089034865038e+00, -7.0195767818990423e-02) & (1.0004601734442544e+00, -1.0831789414602214e-01) \\ 
10 & (-2.1187164479466330e-03, 1.7941441867523995e-03) & (1.0018875176084265e+00, -2.7751848847786847e-02) & (1.0026573677712294e+00, -4.4701214561704387e-02) \\ 
11 & (-5.8905056307767426e-04, 4.6566478562792932e-04) & (1.0005545034546577e+00, -8.9490423218901473e-03) & (1.0011322338597117e+00, -1.6443931400312257e-02) \\ 
12 & (-1.2077673132135951e-04, 1.0098321836765240e-04) & (1.0000451233888372e+00, -1.0868327060711441e-03) & (1.0002029853631049e+00, -4.0783795041548880e-03) \\ 
\bottomrule
\end{tabular}%
}
\label{tab:sqrtcoeffs}
\end{sidewaystable}

\begin{figure}
\centering
\includegraphics[width= .6\textwidth]{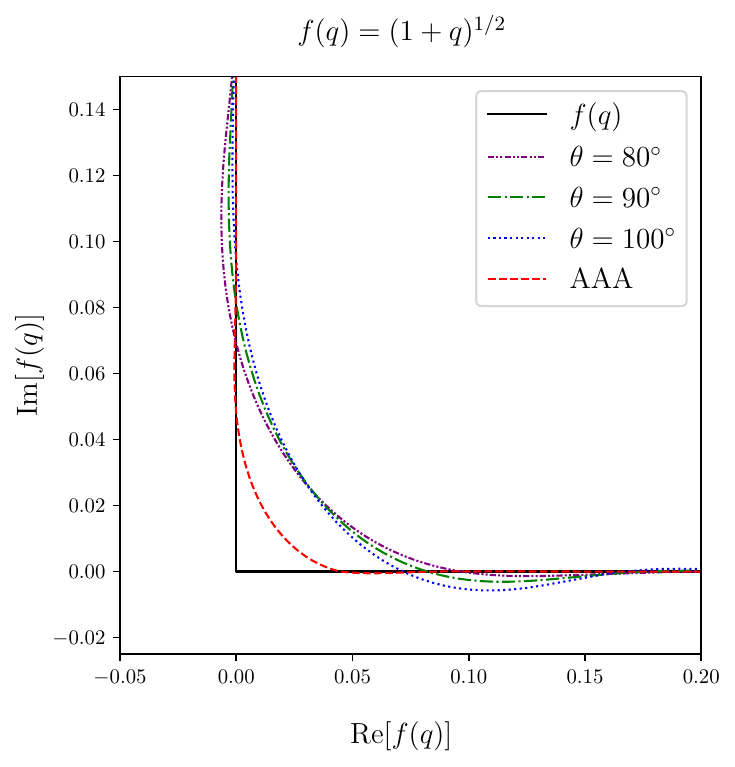}
\caption{The mapping of the real line segment onto the complex plane by the square root operator $\sqrt{1+q}$ zoomed in around the origin.
Shown are the analytic result (solid black), and approximations using the AAA algorithm (dashed red), the ROP method with $\theta=80^\circ$ (dashed-dot-dotted purple), $\theta=90^\circ$ (dash-dotted green),  $\theta=100^\circ$ (dotted blue).
All approximated operators are of order $N=12$.}
\label{fig:sqrt_comp_zoom}
\end{figure}

\begin{figure}
\centering
\includegraphics[width= \textwidth]{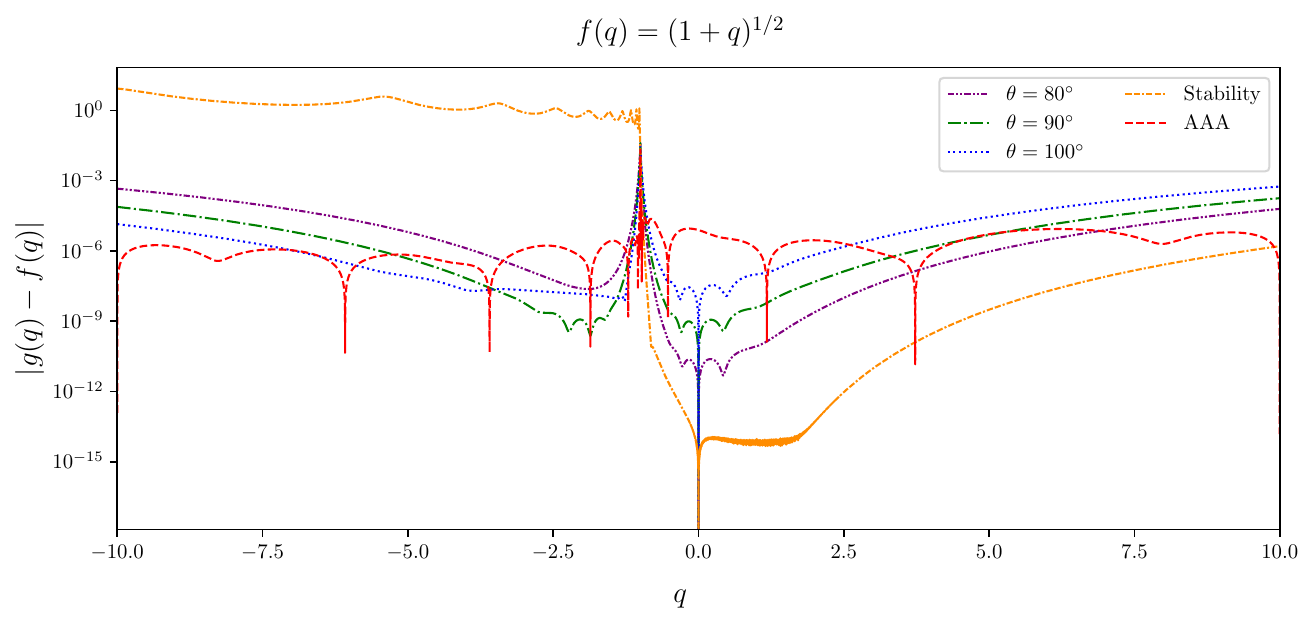}
\caption{The error of the approximations $g(q)$ relative to the analytic function $f(q)=(1+q)^{1/2}$ for $q$ in the range $(-10,10)$.
Shown on are the approximations using the stability constraint method (dash-dash-dotted orange), the AAA algorithm (dashed red), and the ROP method with $\theta=80^\circ$ (dashed-dot-dotted purple), $\theta=90^\circ$ (dash-dotted green),  $\theta=100^\circ$ (dotted blue).
All approximated operators are of order $N=12$.}
\label{fig:sqrt_comp_err}
\end{figure}

\begin{figure*}
\centering
\vspace{-1cm}
\subfloat[]{\includegraphics[width= .38\textwidth]{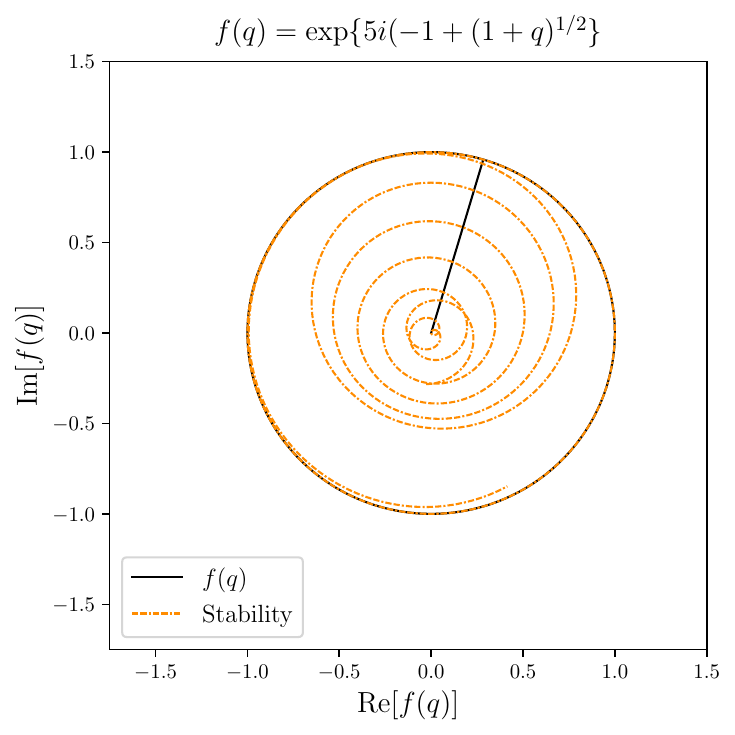}}
\subfloat[]{\includegraphics[width= .38\textwidth]{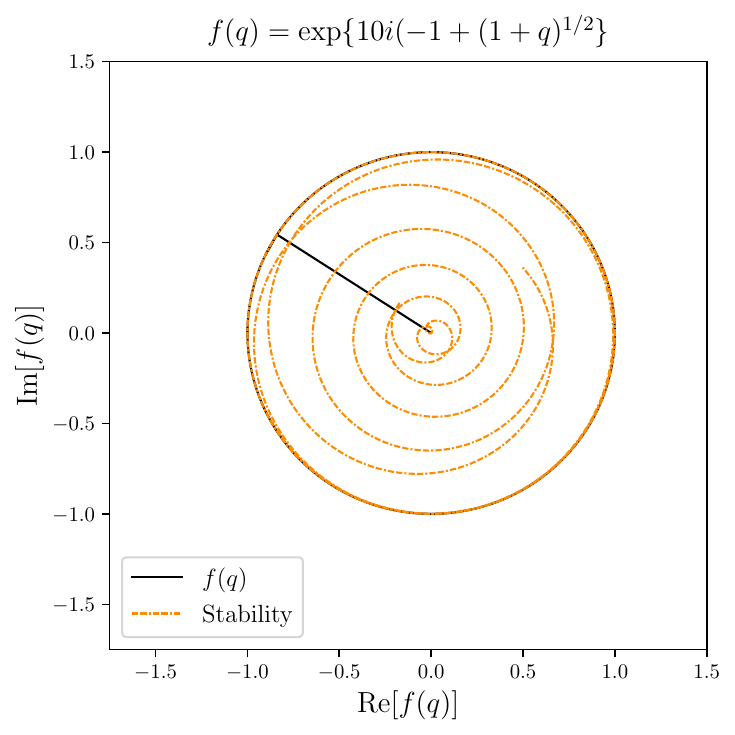}}\\ \vspace{-.2cm}
\subfloat[]{\includegraphics[width= .38\textwidth]{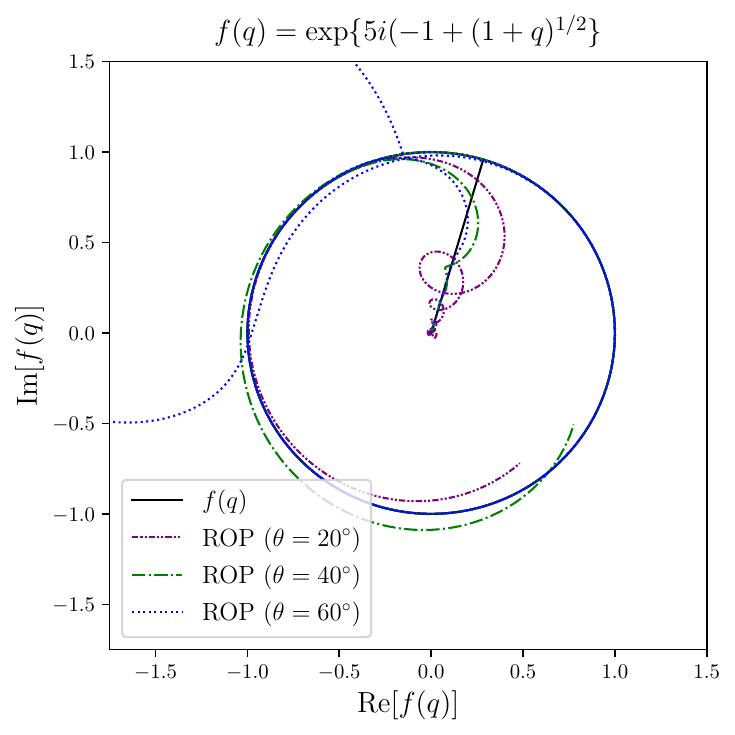}}
\subfloat[]{\includegraphics[width= .38\textwidth]{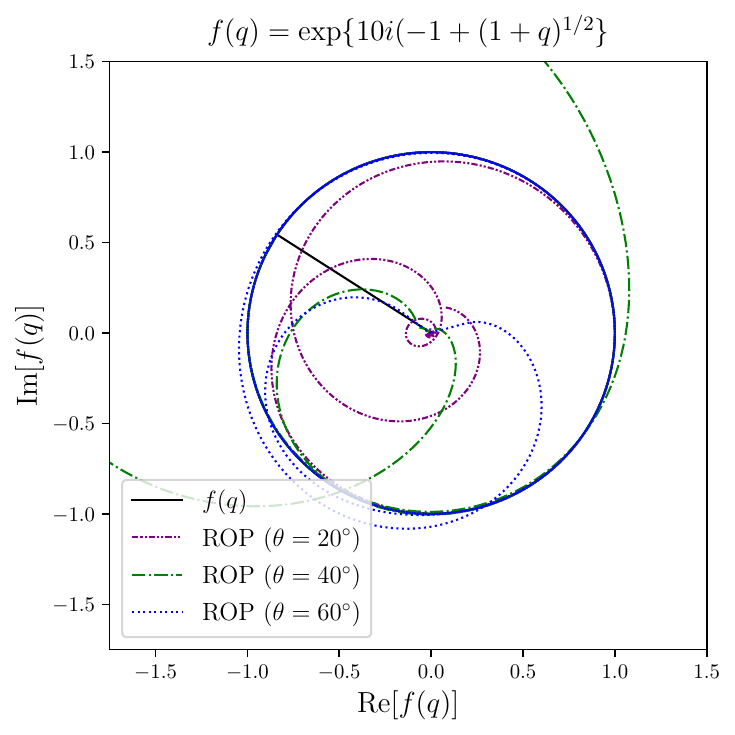}}\\\vspace{-.2cm}
\subfloat[]{\includegraphics[width= .38\textwidth]{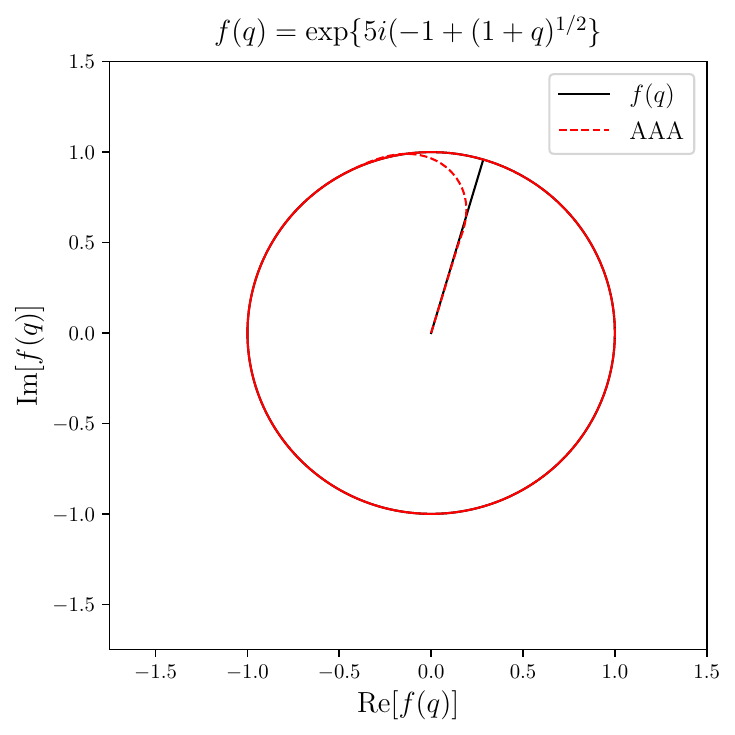}}
\subfloat[]{\includegraphics[width= .38\textwidth]{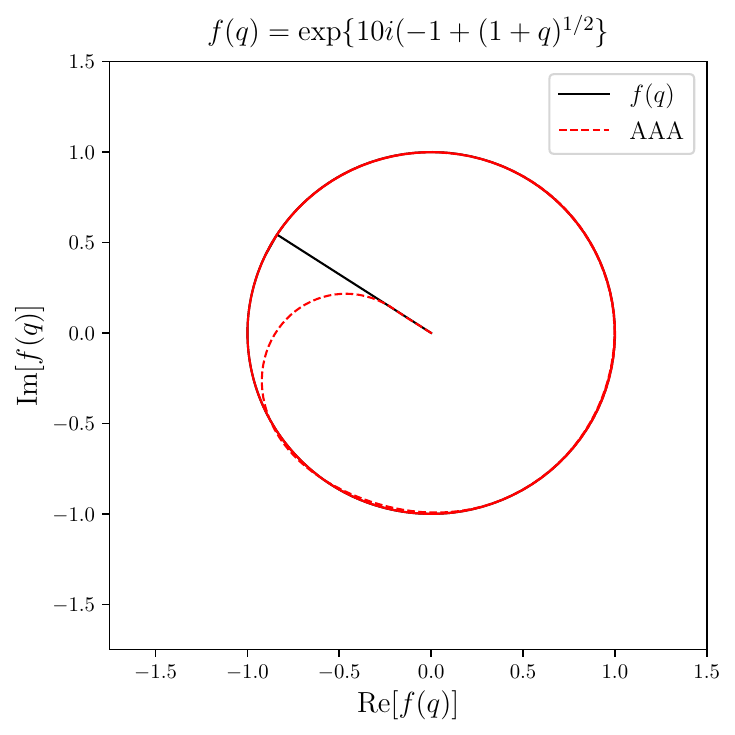}}
\caption{The mapping of the real line segment $(-10,10)$ onto the complex plane by the exponential operator $\exp\{i\sigma(-1+\sqrt{1+q})\}$ with (a),(c),(e) $\sigma=5$ and (b),(d),(f) $\sigma=10$.
Shown (on all panels) are the the analytic result (solid black), 
and approximations using (a),(b) the stability constraint method with $g(-3) = 0$ (dash-dash-dotted orange), 
and (c),(d) the ROP method with $\theta=20^\circ$ (dash-dot-dotted purple), $\theta=40^\circ$ (dash-dotted green), 
and $\theta=60^\circ$ (dotted blue), and (e),(f) the AAA algorithm (dashed red).
All approximated operators are of order $N=12$.
}\label{fig:exp_comp}
\end{figure*}

\begin{figure}
\centering
\subfloat[]{\includegraphics[width= \textwidth]{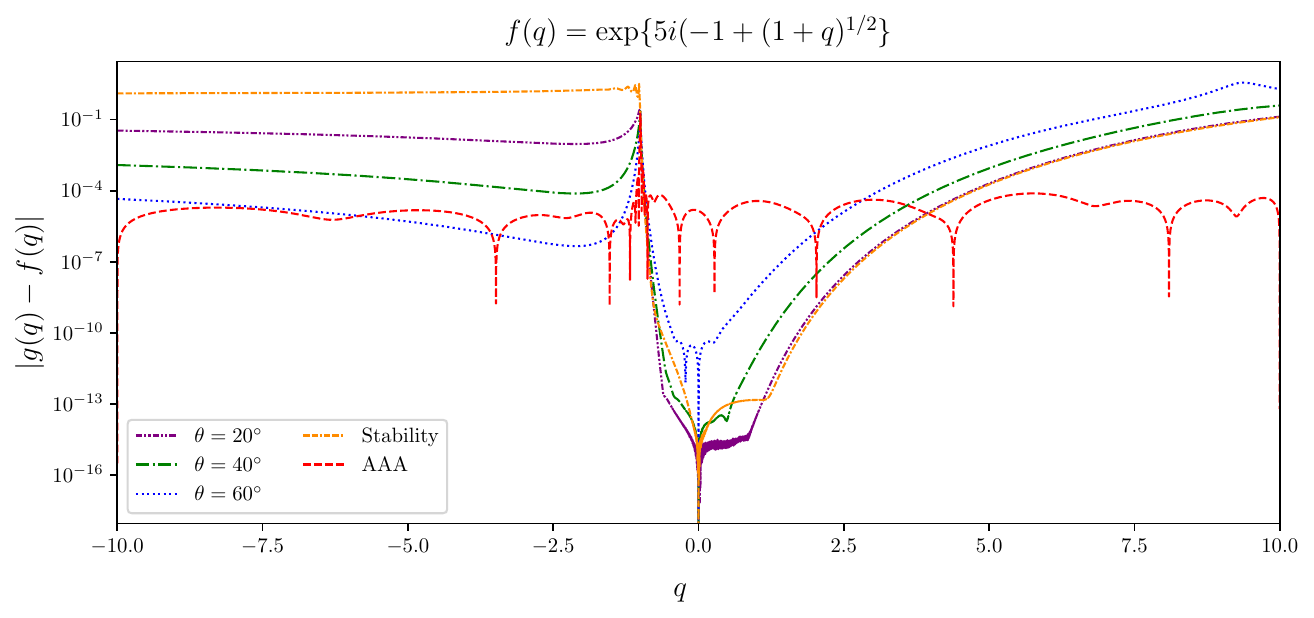}}\\
\subfloat[]{\includegraphics[width= \textwidth]{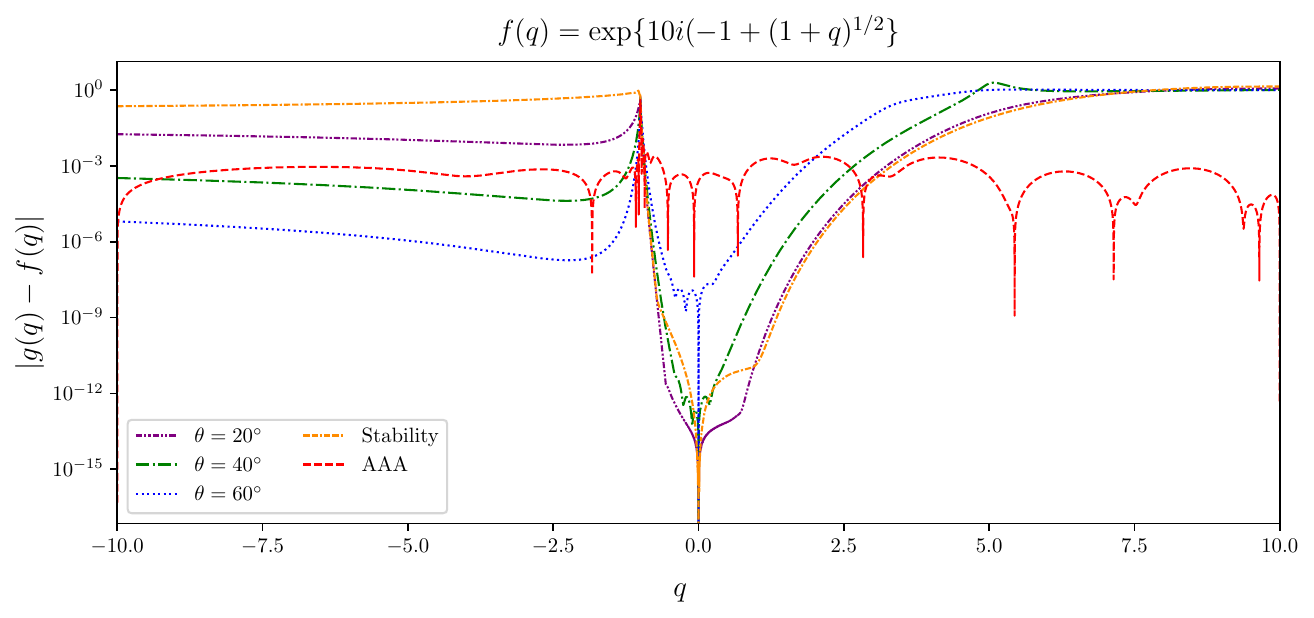}}
\caption{The error of the approximations $g(q)$ relative to the analytic function $f(q)=\exp\{i\sigma(-1+\sqrt{1+q})\}$ with (a) $\sigma=5$ and (b) $\sigma=10$, 
for $q$ in the range $(-10,10)$.
Shown on both panels are the approximations using the stability constraint method with $g(-3) = 0$ (dash-dash-dotted orange), 
the AAA algorithm (dashed red), and the ROP method with $\theta=20^\circ$ (dashed-dot-dotted purple), $\theta=40^\circ$ (dash-dotted green),  $\theta=60^\circ$ (dotted blue).
All approximated operators are of order $N=12$.}
\label{fig:exp_comp_err}
\end{figure}

\begin{figure*}
\centering
\vspace{-1cm}
\subfloat[]{\includegraphics[width= .38\textwidth]{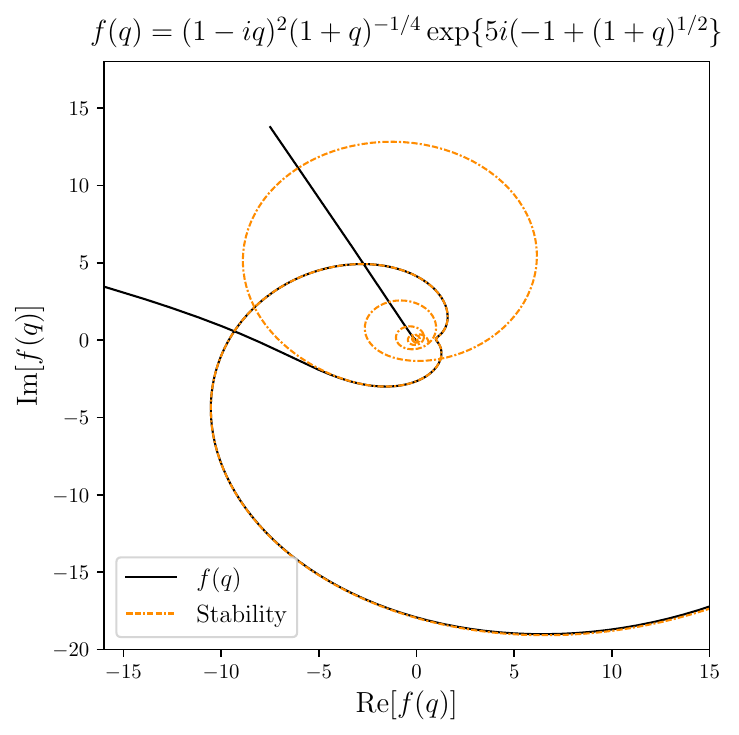}}
\subfloat[]{\includegraphics[width= .38\textwidth]{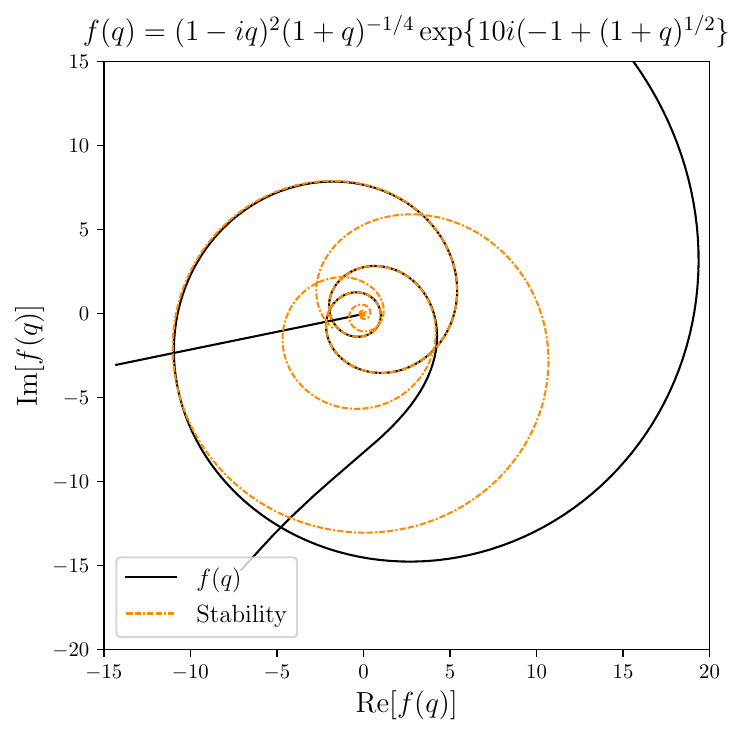}}\\\vspace{-.2cm}
\subfloat[]{\includegraphics[width= .38\textwidth]{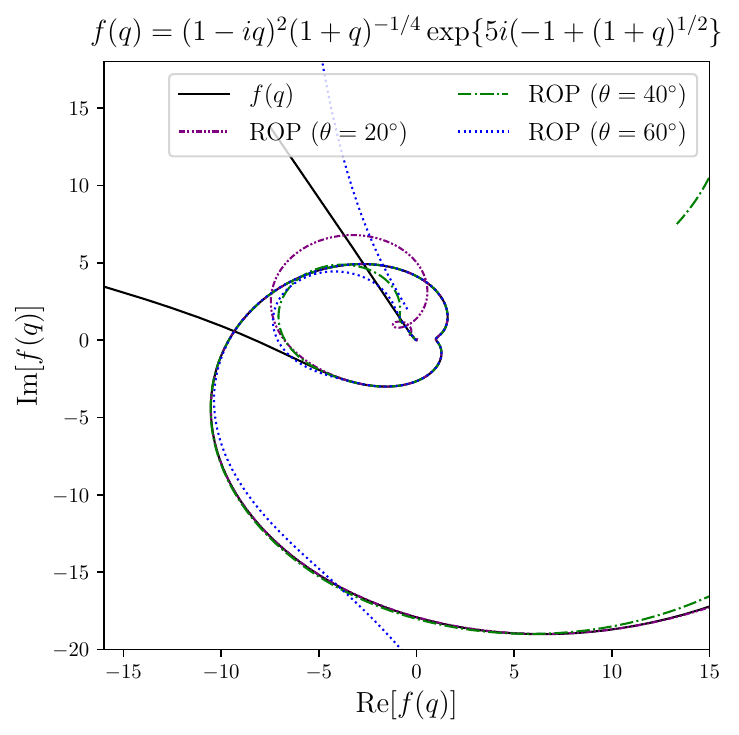}}
\subfloat[]{\includegraphics[width= .38\textwidth]{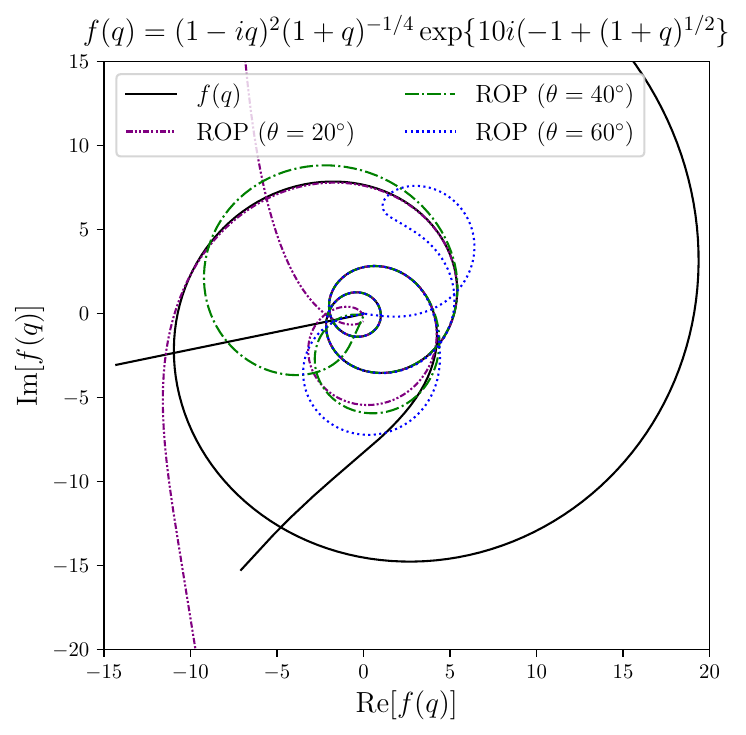}}\\\vspace{-.2cm}
\subfloat[]{\includegraphics[width= .38\textwidth]{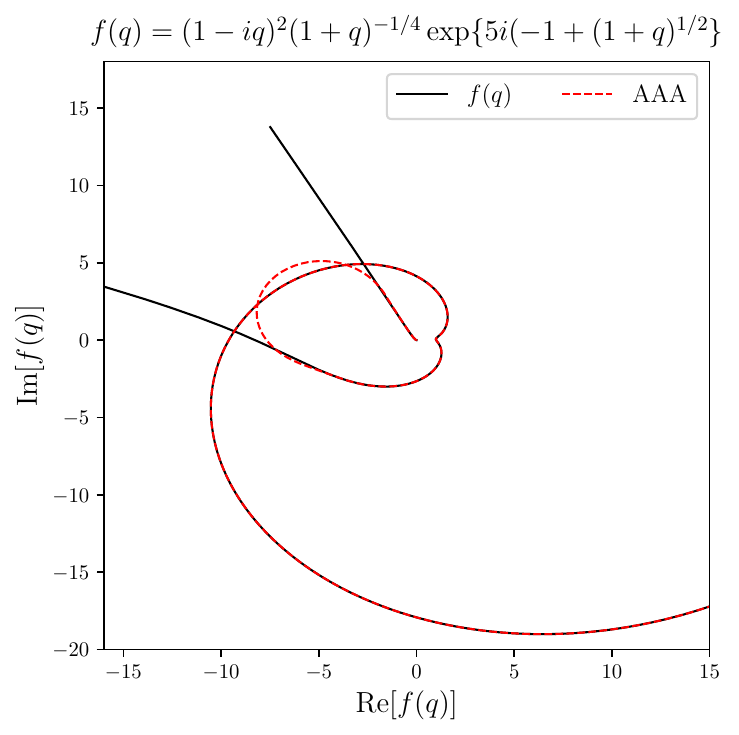}}
\subfloat[]{\includegraphics[width= .38\textwidth]{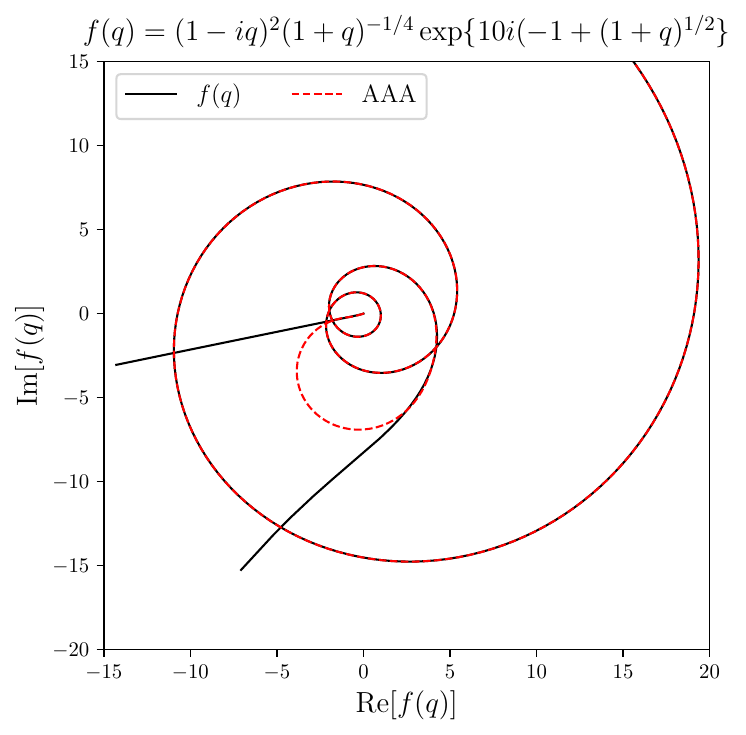}}
\caption{The mapping of the real line segment $(-10,10)$ onto the complex plane by the self-starter operator $(1-iq)^2(1+q)^{-1/4}\exp\{i\sigma(-1+\sqrt{1+q})\}$ with (a),(c),(e) $\sigma=5$ and (b),(d),(f) $\sigma=10$.
Shown are (on all panels) the the analytic result (solid black), and approximations using (a),(b) the stability constraint method (dashed orange),  (c),(d) the ROP method with $\theta=20^\circ$ (dash-dot-dotted purple), $\theta=40^\circ$ (dash-dotted green), and $\theta=60^\circ$ (dotted blue), and (e),(f) AAA algorithm (dashed red).
All approximated operators are of order $N=12$.
}\label{fig:ss_comp}
\end{figure*}

In Fig. \ref{fig:sqrt_comp}, we show the mapping of the real line segment $(-10,10)$ onto the complex plane by the square root operator $f(q) =\sqrt{1+q}$
against a set of approximations of the operator of order $N=12$.
The solid black line is the analytic result; for the approximated operators, 
we show (a) the stability constraint method utilizing a small perturbation off of the real line (dash-dash-dotted orange), 
and (b) the  AAA algorithm with coefficients given in Table \ref{tab:sqrtcoeffs} (dashed red), and the ROP method with $\theta=80^\circ$ (dash-dot-dotted purple), 
$\theta=90^\circ$ (dash-dotted green), $\theta=100^\circ$ (dotted blue).
The evanescent spectrum is not handled accurately by the stability constraint method, although it maps the propagating spectrum with high accuracy.
The AAA and ROP methods perform comparably well in both the evanescent and propagating spectra away from the origin ($f(-1) = 0$).

Fig. \ref{fig:sqrt_comp_zoom} shows a zoomed-in view of the approximations around the origin, 
where, for the ROP approximants, a portion of the real line is mapped onto the lower half of the complex plane;
the deviation below the real line significantly reduced for the approximations found using the AAA algorithm.

Shown in Fig. \ref{fig:sqrt_comp_err} is the error (on a logarithmic scale) of each of the approximations $g(q)$ of the square root operator $f(q)$ for $q$ in the range $(-10, 10)$.
We see that the AAA approximations give relatively consistent error throughout the range of $q$, with dips that correspond to the chosen support points $\{q_j\}$.
The other rational approximations have the greatest accuracy at $q=0$, which is due to the fact that they are derived from Taylor series expansions around that point,
but have errors that increase as $q$ gets further away from the expansion point.
For the ROP method, larger rotations have greater accuracy in the evanescent spectrum at the expense of the accuracy in the propagating spectrum.

In Fig. \ref{fig:exp_comp}, we show the mapping  of the real line segment $(-10,10)$ onto the complex plane by the exponential operator 
$f(q) = \exp\{i\sigma(-1+\sqrt{1+q})\}$ with (a),(c),(e) $\sigma=5$ and (b),(d),(f) $\sigma=10$,
compared with a set of approximations of the operator $g(q)$ of order $N=12$.
The solid black line on all panels is the analytic result; for the approximated operators, 
we show (a),(b) the stability constraint method using the constraint equation $g(-3) = 0$ (dash-dash-dotted orange), 
(c),(d) the ROP method with $\theta=20^\circ$ (dash-dot-dotted purple), $\theta=40^\circ$ (dash-dotted green), and $\theta=60^\circ$ (dotted blue), 
and (e),(f) the AAA algorithm (dashed red).

As with the square root operator, the stability constraint method accurately handles the propagating portion of the spectrum, but does not map the evanescent spectrum properly.
The ROP method is most accurate for the propagating spectrum when the rotation angle $\theta$ is small.
Above $\theta=60^\circ$, the approximation deviates outside of the unit circle even for small values of $\sigma$; propagating eigenvalues are mapped outside of the unit circle, meaning that their associated modes grow exponentially.
As $\sigma$ is increased, the upper limit of $\theta$ for which ROP approximation does not exceed the unit circle is decreased;
$\theta$ must sufficiently small, which reduces the accuracy of the approximation for values $q < -1$.
For both the stability constraint and ROP methods, with $\sigma=10$, the mapping begins to deviate within the unit circle as $q\rightarrow10$.

The approximation from the AAA algorithm is accurate for the propagating spectrum 
-- it has none of the previously described issues, such as mapping values outside of the unit circle, over a larger range of $\sigma$ --
while mapping the evanescent spectrum with accuracy.

In Fig. \ref{fig:exp_comp_err}, we show the error of each approximation of the exponential operator 
$\exp\{i\sigma(-1+\sqrt{1+q})\}$ with (a) $\sigma=5$ and (b) $\sigma=10$ for $q\in(-10,10)$.
Similar to the case of the square root operator, the AAA algorithm yields approximations that have relatively constant error for all $q$, with dips at the support points,
and performs equally well in the evanescent and propagating portions of the spectrum.
The ROP and stability constraint methods give high accuracy around $q=0$, which then increases as $q\rightarrow10$. 
Larger rotations give larger error in the propagating spectrum while being more accurate in the evanescent spectrum, as was the case with the square root operator.
The $\theta=20^\circ$ approximation is relatively comparable in error and behavior in the 
propagating spectrum to the approximation yielded  by the stability constraint method, while handling the evanescent spectrum with more accuracy.

As a result of the sensitivity to $\theta$, the ROP method for approximating the exponential operator is not suitable for most typical use-cases.
Accuracy in the propagating spectrum with large range steps requires small values of $\theta$, which reduces the accuracy of the 
operator applied to the non-propagating portion of the spectrum, causing stability issues.
Similarly, the stability constraint method does not yield operators that are stable for all elastic and fluid-elastic cases, 
even when using multiple stability constraint equations, as the evanescent spectrum is not sufficiently annihilated.

In Fig. \ref{fig:ss_comp}, we show the mapping  of the real line segment $(-10,10)$ onto the complex plane by the point-source self-starter operator 
$(1-iq)^2(1+q)^{-1/4}\exp\{i\sigma(-1+\sqrt{1+q})\}$ with (a),(c),(e) $\sigma=5$ and (b),(d),(f) $\sigma=10$,
compared with a set of approximations of the operator of order $N=12$.
The solid black line on all panels is the analytic result; for the approximated operators, we show (a),(b) the stability constraint method with constraint equation $g(-3) = 0$,
(c),(d) the ROP method with $\theta=20^\circ$ (dash-dot-dotted purple), $\theta=40^\circ$ (dash-dotted green), and $\theta=60^\circ$ (dotted blue), and (e),(f)  the AAA algorithm (dashed red).

The self-starter operator is unique amongst the PE operators in that the analytic solution goes to infinity as $q\rightarrow-1$. 
This can be seen in the figure with the analytic function (black line) leaving and returning to the plot area.
Approximations generally treat this region as a smooth curve, and do not necessarily accurately capture the mapping of all wave numbers, 
especially in the region around $q=-1$.

The stability constraint method does quite well in mapping the section of the operator $q > -1$ for $\sigma=5$, but is not as accurate in the $q < -1$ region 
(the ``leg'' of the analytic function that approaches the origin).
For $\sigma=10$ the approximation deviates fairly significantly from the analytic function as $q\rightarrow10$.
The ROP method is most accurate when the rotation angle $\theta$ is small.
Small rotations, however, do a poorer job of capturing the region around $q=-1$.
Above $\theta=60^\circ$, the approximation begins to deviate from the analytic function even for small $\sigma$.
For larger values of $\sigma$, small values of $\theta$ yield decent approximations, though they deviate from the analytic function as $q\rightarrow10$, 
but for larger $\theta$, the ROP approximation breaks down entirely.
These behaviors are similar to those for the exponential operator.

The AAA approximation maintains good accuracy for the entire mapping, and performs well regardless of value of $\sigma$.
An accurate starting field is essential for simulations to yield correct results, 
and while the stability constraint method for approximation of the self-starter operator
has yielded excellent results, the approximations using the AAA algorithm are significantly more accurate.

\section{Examples and discussion}

Many studies have demonstrated the ability of parabolic equation methods to accurately model wave propagation in laterally-varying
coupled fluid-elastic waveguides\cite{collins2015,collins2015.1,woolfe2016,collins2021,collins2021.2}.
We will focus on where the standard operator approximations break down in either stability or accuracy, 
and see if the approximations from the AAA algorithm perform better.

In the examples that follow, we will use the same seismoacoustic PE formulation as in those benchmarking studies;
fluid-fluid and solid-solid vertical interfaces are handled using a single-scattering approach\cite{collins2012},
while fluid-solid vertical interfaces are subject conservation of energy for the compressional wave and vanishing tangential stress\cite{collins2019.2}.
We take $N=12$ for all rational approximations. 

All simulations applying the AAA algorithm to approximate the square root operator use the coefficients found in Table \ref{tab:sqrtcoeffs},
and use the AAA approximation for the self-starter operator.
Simulations using the ROP method to approximate the square root use the stability constraint method for the self-starter, 
with constraint equation $g(-3) = 0$.

In order to minimize errors due to the discretization and isolate the behavior of the different rational approximations, 
the grid spacing for simulations using the square root operator 
was chosen to be $\Delta r = \lambda_0/16,\, \Delta z = \lambda_0/64$, with the reference wavelength $\lambda_0 = c_0/f$, $f$ the frequency in Hz,
 and reference speed $c_0=2000$ m/s.
For simulations using the exponential operator, we use $\Delta r=1.5\lambda_0$ and $\Delta z = \lambda_0/64$.
Benchmark solutions were computed using the COMSOL Multiphysics software suite\cite{comsol}.

\begin{figure}
\centering
\includegraphics[width=\textwidth]{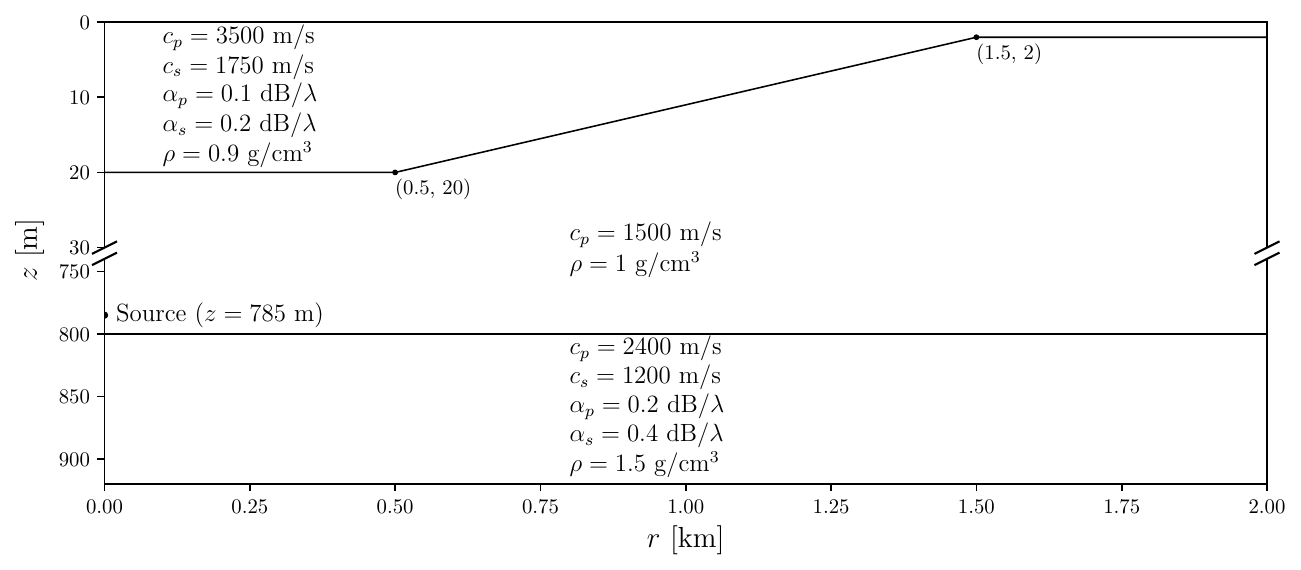}
\caption{Schematic of the waveguide for Example A, which involves a tapering thin elastic layer overlaying a fluid medium and an elastic halfspace.}\label{fig:exAwg}
\end{figure}

\begin{figure}
\centering
\includegraphics[width=\textwidth]{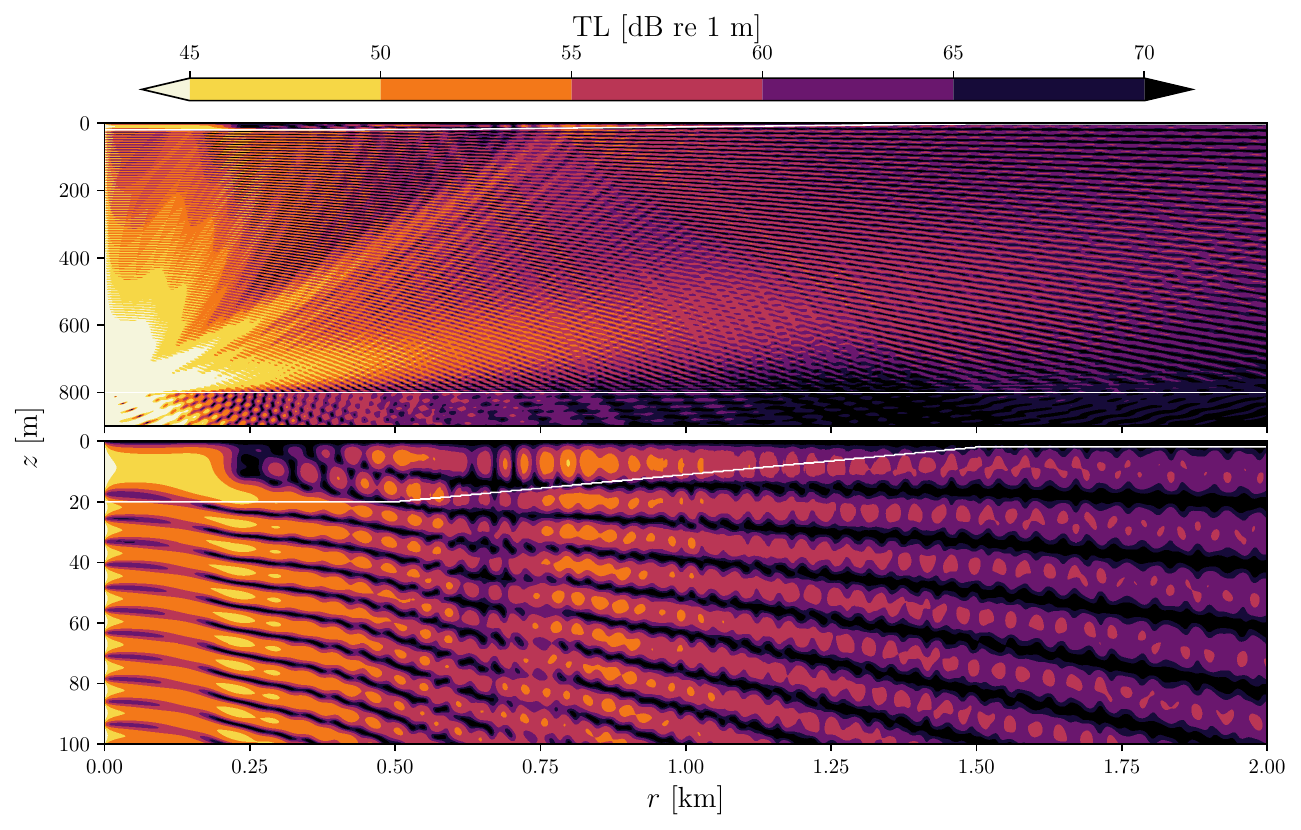}
\caption{2D transmission loss (TL) plots for Example A computed with the PE using the square root operator approximated by the AAA algorithm. 
The bottom panel shows a zoomed-in section at the top of the computational domain.}\label{fig:exAclr}
\end{figure}

We consider the field produced by a compressional point source in an azimuthally symmetric waveguide parameterized by coordinates $(r,z)$.
In what follows, the transmission loss (TL), including cylindrical spreading, in an elastic medium is defined as 
\eq{
\mathrm{TL}_\mathrm{elastic} = -20 \log_{10}\left(\sigma_{zz}\right) + 10 \log_{10}(r)~\text{ [dB re 1 m]}\,, \nn
}
with $\sigma_{zz}$ the $zz$ component of the stress tensor.
In the fluid media,
\eq{
\mathrm{TL}_\mathrm{fluid} = -20 \log_{10}(p) + 10 \log_{10}(r)~~~~\text{ [dB re 1 m]}\,, \nn
}
with $p$ the pressure.

\begin{figure}
\centering
\subfloat[]{\includegraphics[width=.97\textwidth]{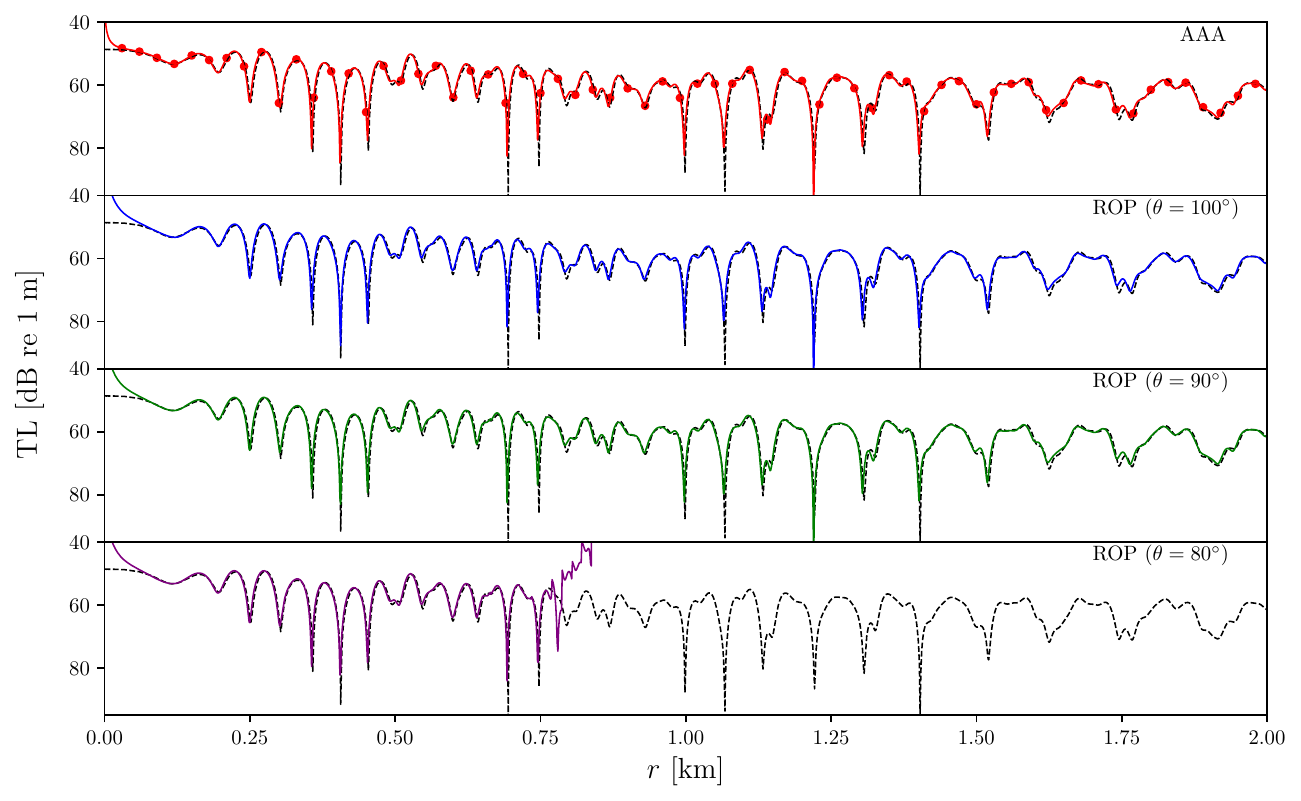}}\\
\subfloat[]{\includegraphics[width=.97\textwidth]{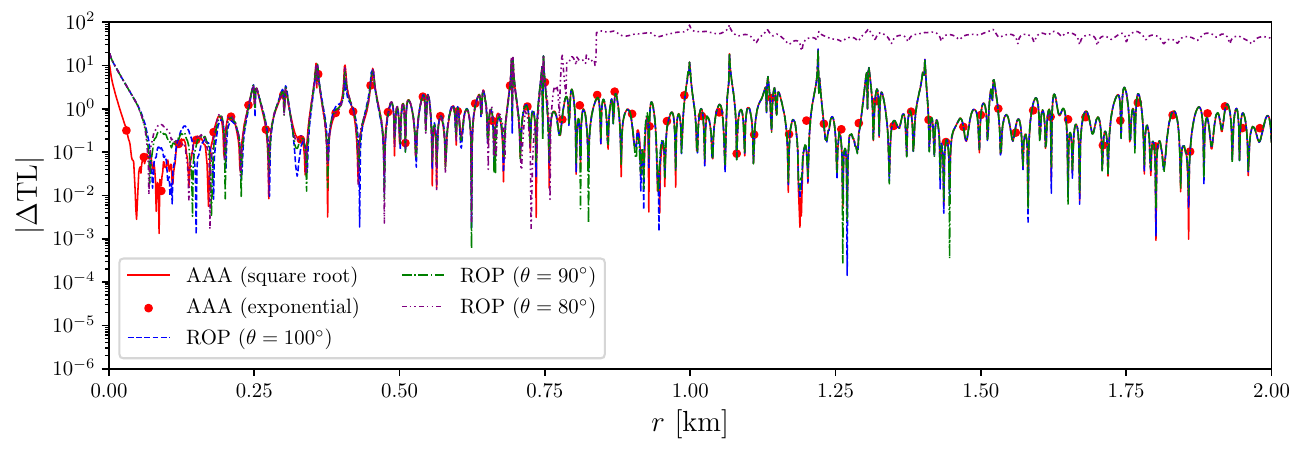}}
\caption{(a) Transmission loss (TL) at depth $z = 300$ m for Example A. 
The benchmark solution computed using the finite-element method is displayed in each panel as a black dashed line.
From top to bottom, the panels are for the AAA approximation (solid red line for the square root operator, red circles for the exponential operator with $\Delta r = 30$ m),
and the ROP method for the square root operator with $\theta=100^\circ$ (solid blue), $\theta=90^\circ$ (solid green), and $\theta=80^\circ$ (solid purple).
(b) The absolute TL error for each approximation method, given by difference between the PE results and the benchmark solution.}\label{fig:exATL}
\end{figure}

The environment for Example A is shown in Fig. \ref{fig:exAwg}. 
From top to bottom, the waveguide consists of a thin layer of elastic material, with density $\rho = 0.9$ g/cm$^3$, compressional wave speed $c_p=3500$ m/s, 
shear wave speed $c_s = 1750$ m/s, compressional attenuation $\alpha_p$ = 0.1 dB/$\lambda$, and shear attenuation $\alpha_s$ = 0.2 dB/$\lambda$, with $\lambda$ the wavelength.
This layer has thickness 20 m for $r < 500$ m and 2 m for $r > 1500$ m, with a linear decrease in thickness from 500 m to 1500 m range.
This thin elastic layer overlays a fluid layer that extends to $z= 800$ m, with density $\rho = 1.0$ g/cm$^3$ and sound speed $c_p = 1500$ m/s.
The lower portion of the waveguide consists of an elastic halfspace with density $\rho = 1.5$ g/cm$^3$, compressional wave speed $c_p=2400$ m/s, 
shear wave speed $c_s = 1200$ m/s, compressional attenuation $\alpha_p$ = 0.2 dB/$\lambda$, and shear attenuation $\alpha_s$ = 0.4 dB/$\lambda$.
The domain is truncated at the bottom with a perfectly-matched layer. 
A 100 Hz source is placed at $z = 785$ m, and the field is propagated to 2 km range.

Fig. \ref{fig:exAclr} shows the 2D TL for Example A computed using the square root operator approximated by the AAA algorithm. 
The lower panel is a zoomed-in section of the domain showing the variation of the upper elastic layer.

Fig. \ref{fig:exATL}(a) shows the TL curves for Example A at $z=300$ m for a set of approximations of the PE operators.
The results using the AAA algorithm are shown in the upper panel for both the square root operator (solid blue line)
and the exponential operator with $\Delta r = 30$ m (red circles).
The results using the square root operator (solid blue lines) approximated using the ROP method with $\theta=100^\circ,90^\circ,$ and $80^\circ$ are shown on the subsequent panels.
The absolute errors in the TL for each approximation method as a function of range are shown in Fig. \ref{fig:exATL}(b), where $\Delta \mathrm{TL} \equiv \mathrm{TL}_\mathrm{PE} - \mathrm{TL}_\mathrm{FEM}$.

For this example, the ROP method of approximating the square root operator with $\theta<90^\circ$ breaks down in the laterally-varying region of the domain.
The magnitude of the error using the AAA approximation (both square root and exponential forms of the operator) is significantly less at short range, due to the more accurate representation of the self-starter operator.
The errors using the AAA approximation in the far field are comparable to those using the ROP approximation method with appropriate rotation angles $\theta\geq90$.

Notably, the split-step Pad\'e approach using AAA algorithm approximation of the exponential has excellent agreement with the benchmark solutions. 
This addresses a known shortcoming of approximations as applied to certain fluid-elastic waveguides; 
with the improved approximations of the exponential operator, wave propagation in these waveguides can be 
simulated with a much larger range step, while maintaining accuracy and stability.

\begin{figure}
\centering
\includegraphics[width=\textwidth]{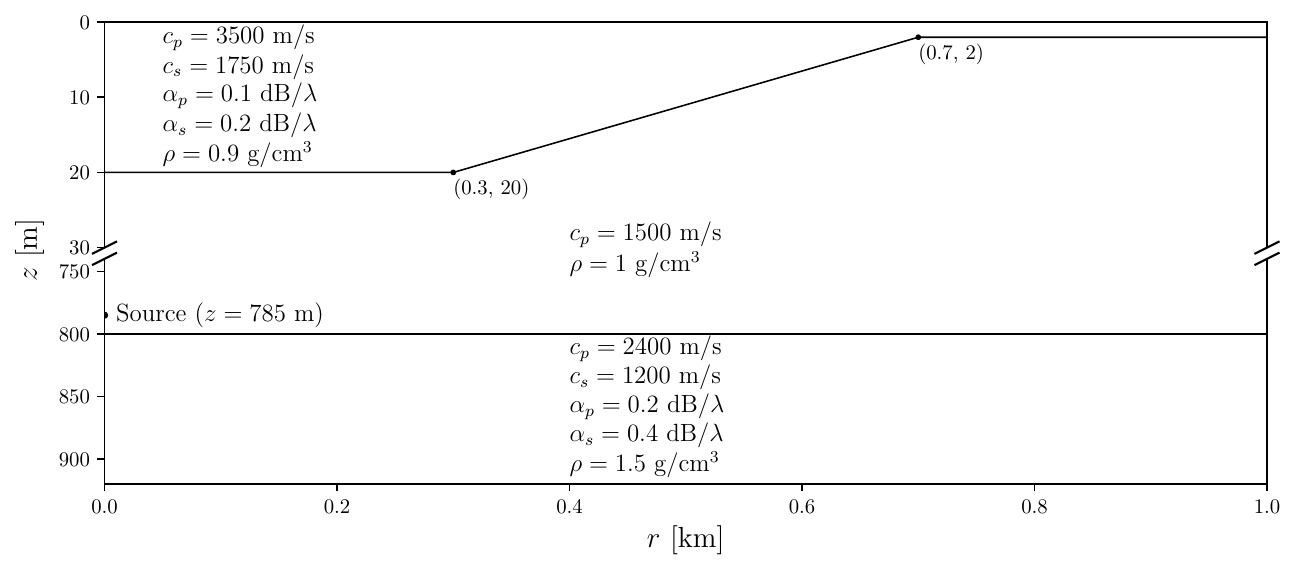}
\caption{Schematic of the waveguide for Example B, which involves a tapering thin elastic layer overlaying a fluid medium and an elastic halfspace.}\label{fig:exBwg}
\end{figure}

\begin{figure}
\centering
\includegraphics[width=\textwidth]{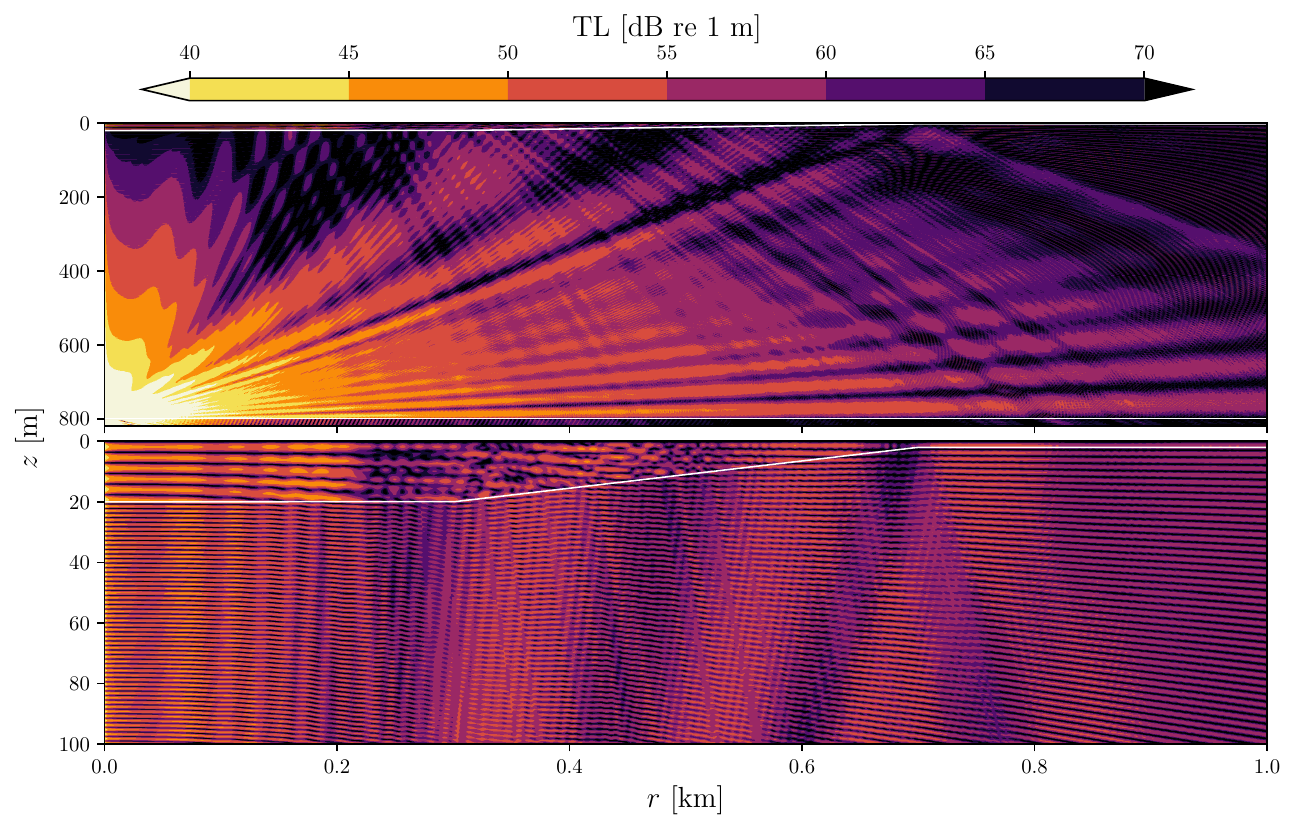}
\caption{2D transmission loss (TL) plots for Example B computed with the PE using the square root operator approximated by the AAA algorithm. 
The bottom panel shows a zoomed-in section at the top of the computational domain.}\label{fig:exBclr}
\end{figure}

\begin{figure}
\centering
\includegraphics[width=\textwidth]{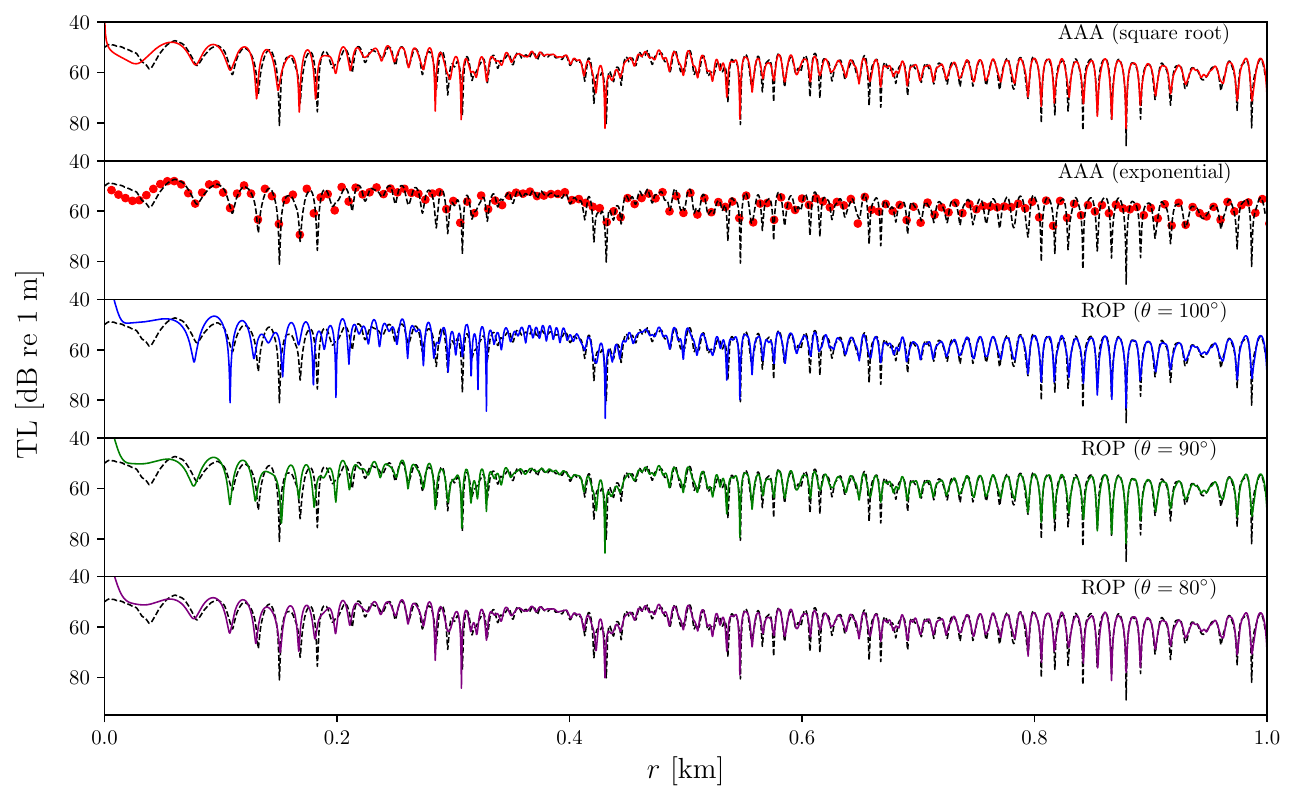}
\caption{Transmission loss (TL) at depth $z = 300$ m for Example B. 
The benchmark solution computed using the finite-element method is displayed in each panel as a black dashed line.
From top to bottom, the panels are for the AAA approximation of the square root operator (red solid), AAA approximation of the exponential operator with $\Delta r = 6$ m (red circles),
and the ROP method for the square root operator with $\theta=100^\circ$ (blue solid), $\theta=90^\circ$ (green solid), and $\theta=80^\circ$ (purple solid). }\label{fig:exBTL}
\end{figure}

The waveguide for Example B is shown in Fig. \ref{fig:exBwg}.
The layer properties are identical to that of Example A; the difference between the examples is in the tapering of the upper thin elastic layer, 
which has thickness 20 m for $r < 300$ m and 2 m for $r > 700$ m, with a linear decrease in thickness from 300 m to 700 m range.
A 500 Hz source is placed at $z = 785$ m, and propagation is limited to 1 km range to allow for computation of benchmark solutions
using the finite-element method.

Fig. \ref{fig:exBclr} shows the 2D TL for Example B computed using the square root operator approximated by the AAA algorithm. 
The lower panel is a zoomed-in section of the domain showing the variation in thickness of the upper elastic layer.
Fig. \ref{fig:exBTL} shows the TL curves for Example B at $z=300$ m for a set of approximations of the PE operators.
The results using the AAA algorithm are shown in the upper panel the square root operator (solid red line)
and second panel for the exponential operator with $\Delta r = 5$ m (red circles).
The results using the square root operator approximated using the ROP method with $\theta=100^\circ$ (solid blue),
$90^\circ$ (solid green),  and $80^\circ$ (solid purple) are shown on the subsequent panels.

As with Example A, simulations using the AAA algorithm for approximations give excellent agreement with the benchmark solution for Example B, 
including when taking range steps of 6 m using the exponential operator.
For the ROP approximation of the square root operator with $\theta=90^\circ$ and $100^\circ$, at ranges where the top layer thickness is 20 m ($r < 400$ m), 
there are deviations from the benchmark result.
These deviations are possibly due to a mishandling of propagating and evanescent modes in the 20 m thick ice layer.
The errors are reduced as the top layer thickness tapers, presumably due to mode cutoff; 
the different approximations have relatively comparable errors for $r>600$ m.

Unlike in Example A,  the $\theta=80^\circ$ approximation does not completely break down and, 
of the three rotations tested, has the best agreement with the reference solution, though the error is still larger than that of the simulations using the AAA algorithm.
This leads to the observation that the optimal value of $\theta$ is dependent on frequency and/or waveguide parameters.

To quantify this, we present two examples. 
We define the average error in the transmission loss per unit length (in dB) as 
\eq{
E(\theta,f,z) = \left( \frac{1}{r_\mathrm{max} - r_\mathrm{min}}\right) \times \int_{r_\mathrm{min}}^{r_\mathrm{max}} dr \left|\mathrm{TL}_\mathrm{PE}(\theta,f,z) - \mathrm{TL}_\mathrm{FEM}(f,z)\right|\,.
}
The same metric can be calculated for the approximation of the square root from the AAA algorithm.
For all that follows, we take $r_\mathrm{min}=25$ m.

\begin{figure}
\centering
\includegraphics[width=\textwidth]{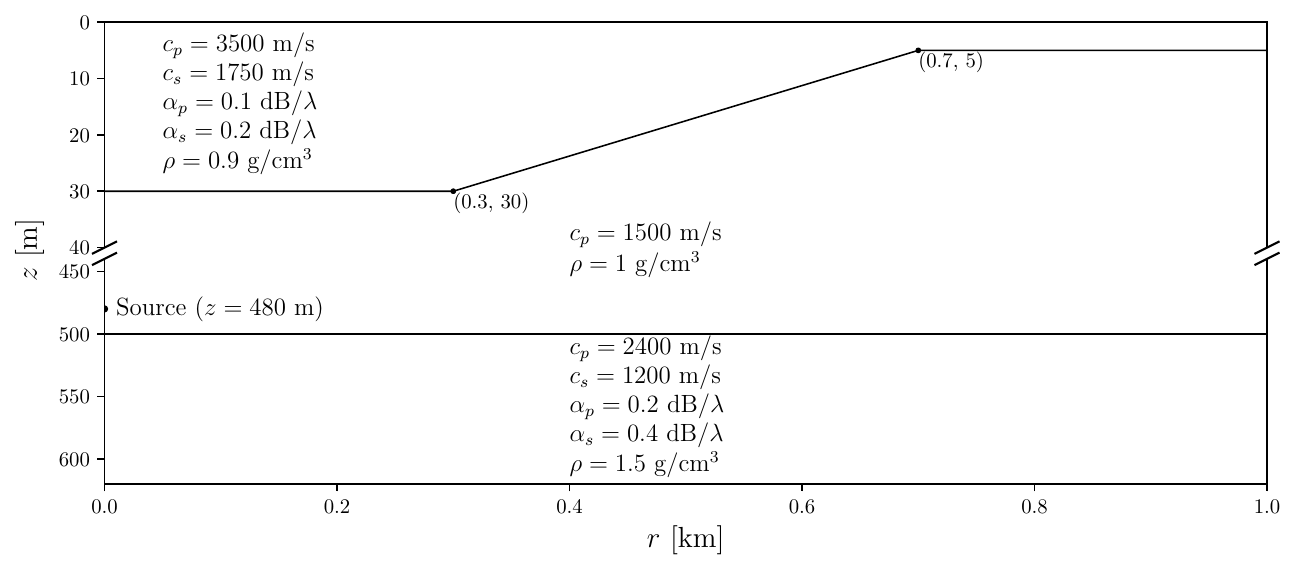}
\caption{Schematic of the waveguide for Example C, which involves a tapering thin elastic layer overlaying a fluid medium and an elastic halfspace.}\label{fig:exCwg}
\end{figure}

\begin{figure}
\centering
\includegraphics[width=\textwidth]{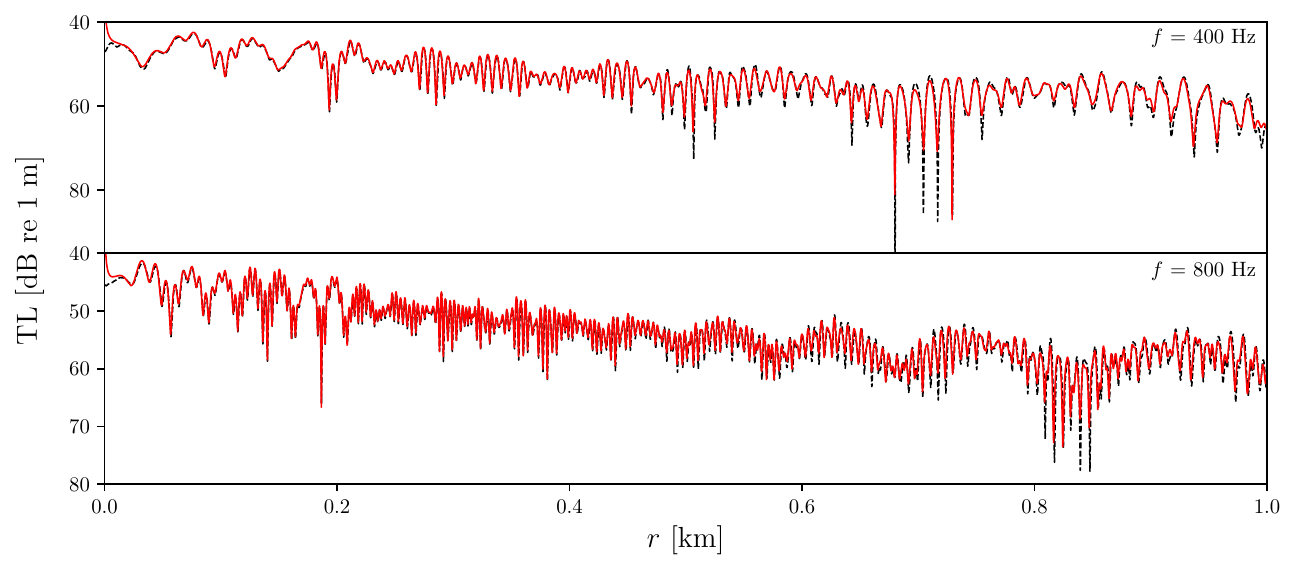}
\caption{Transmission loss (TL) at depth $z = 300$ m for Example C at (upper panel) $f=400$ Hz  and (lower panel) $f=800$ Hz.
The solid red curves are results from simulations using the AAA algorithm to approximate the square root. 
The benchmark solution computed using the finite-element method is displayed in each panel as a black dashed line.
}\label{fig:exCTL}
\end{figure}

\begin{figure}
\centering
\includegraphics[width=.7\textwidth]{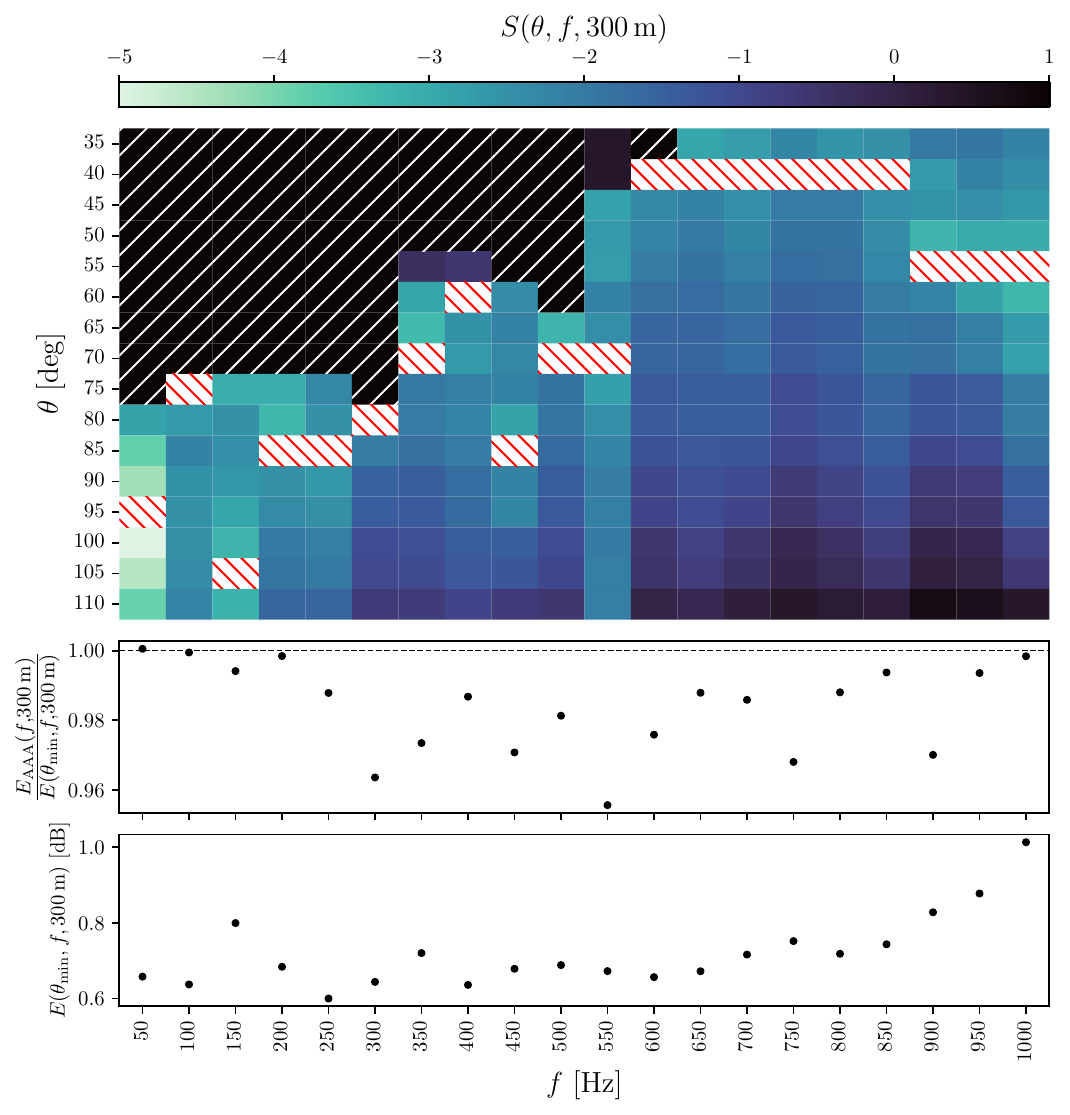}
\caption{Errors in transmission loss at $z=300$ m for Example C using approximations computed using 
(upper panel) the ROP method as a function of rotation angle $\theta$ and frequency $f$ relative to the optimal ROP $\theta$ value, and 
(middle panel) the AAA algorithm relative to the optimal ROP $\theta$ value. 
The lower panel shows the average error in transmission loss in dB using the ROP approximation with the optimal $\theta$ value.
The black squares with white hatches indicate combinations of $\theta$ and $f$ for which the simulation results diverged, while
the white squares with red hatches indicate the optimal value of $\theta$ for each frequency.
See text for more detail.}
\label{fig:thetaswpC}
\end{figure}

The waveguide for Example C is shown in Fig. \ref{fig:exCwg}, and the properties of the three layers (elastic upper, fluid, and elastic bottom) are identical to those
in Examples A and B.
The fluid layer extends to 500 m, is overlayed by a thin layer of elastic material with thickness 30 m for $r < 300$ m and 5 m for $r > 700$ m, 
with a linear decrease in thickness from 300 m to 700 m range.
A source is placed at $z = 480$ m and the field is propagated out to 1 km range.

TL curves at $z=300$ m for frequencies 400 and 800 Hz for Example C are shown in Fig. \ref{fig:exCTL} 
for simulations using the AAA algorithm for approximating the square root.
The AAA algorithm yields operators that give excellent agreement with the benchmark curves across all frequencies studied. 

The upper panel of Fig. \ref{fig:thetaswpC} shows the results of the sweep over values of $\theta$ from $35^\circ$ to $110^\circ$ in increments of 5 degrees, 
and frequencies $50 - 1000$ Hz in 50 Hz increments. 
In particular, we show the quantity
\eq{
S(\theta,f,z) = \log_{10}\left[E(\theta,f,z)/E(\theta_\mathrm{min},f,z) - 1\right]\,,
}
 for $z=300$ m depth. 
$S$ characterizes the error for a particular $\theta$ relative to the rotation $\theta_\mathrm{min}$ that minimizes $E$ for a particular frequency $f$ at depth $z$.

The black squares with white hatches indicate combinations of $\theta$ and $f$ for which the simulation results diverged, which we define as $E > 10$ dB/m.
The white squares with red hatches indicate the $\theta_\mathrm{min}$ for each frequency.
At low frequencies, low values of $\theta$ cause simulations to diverge and best accuracy relative to benchmark solutions is obtained with large rotations.
At high frequencies, on the other hand, the error is large with greater rotations, and the best accuracy is obtained if theta is much smaller than 90 degrees.
There is no obvious functional trend for the value of $\theta$ that minimizes the error at any given frequency, other than that $\theta$ must be reduced
as frequency is increased.

The middle panel of Fig. \ref{fig:thetaswpC} shows the error of simulations using the square root operator approximated by the AAA algorithm normalized to the
error using the ROP method with $\theta_\mathrm{min}$ at each frequency.
The lower panel shows the average error in transmission loss using the ROP method to approximate the square root with $\theta_\mathrm{min}$ at each frequency.
There are minor phasing errors that appear as frequency is increased, which causes the overall rise in average error with frequency for all simulations.

The AAA approximation performs as well, if not better, for all frequencies against the most accurate simulations from the $\theta$ values sampled.
We should note that, by doing a finer sweep of $\theta$, it is possible that comparable error values could be obtained; 
this, however, is impractical for most typical use-cases of PE simulations.
In addition, to verify stability, we simulated propagation in this waveguide at 2000 and 2500 Hz, 
and results were stable for approximations found using the AAA algorithm and small values of the rotation angle using the ROP method,
but could not compute benchmark curves using the finite-element method due to computational limitations.

\begin{figure}
\centering
\includegraphics[width=\textwidth]{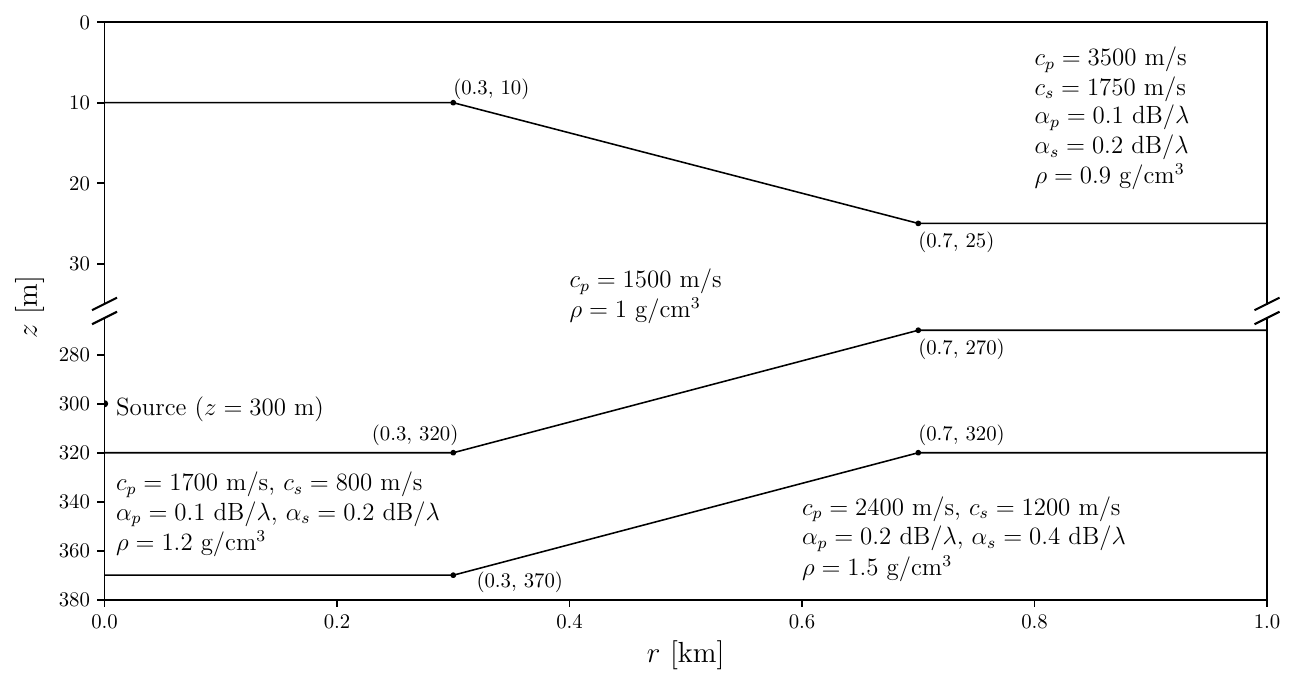}
\caption{Schematic of the waveguide for Example D, which involves a thin elastic layer of increasing thickness overlaying a fluid medium and a two-layered elastic bottom.}\label{fig:exDwg}
\end{figure}

\begin{figure}
\centering
\subfloat[]{\includegraphics[width=\textwidth]{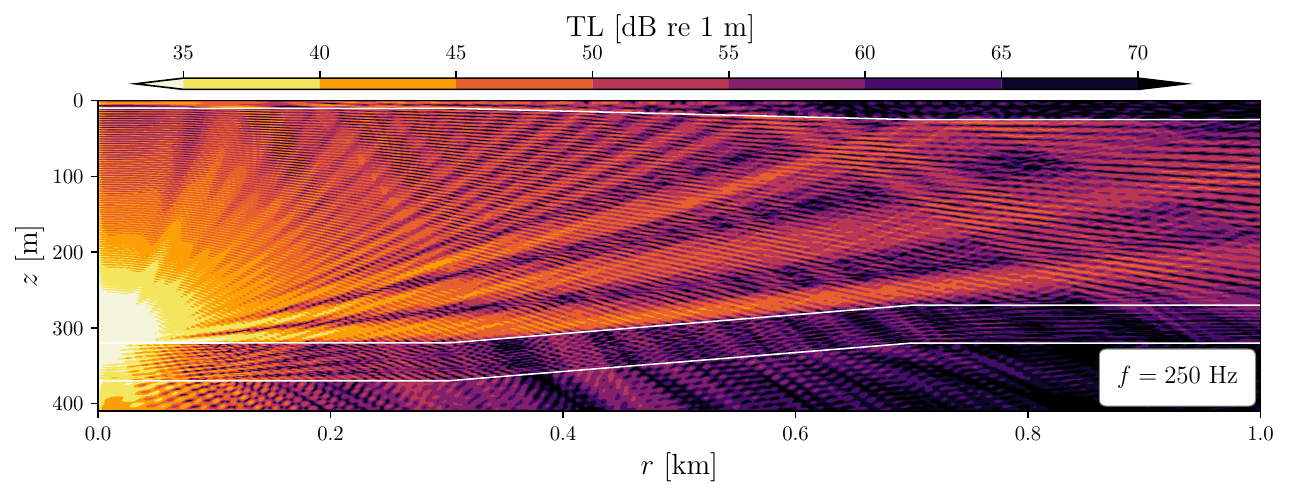}}\\
\subfloat[]{\includegraphics[width=\textwidth]{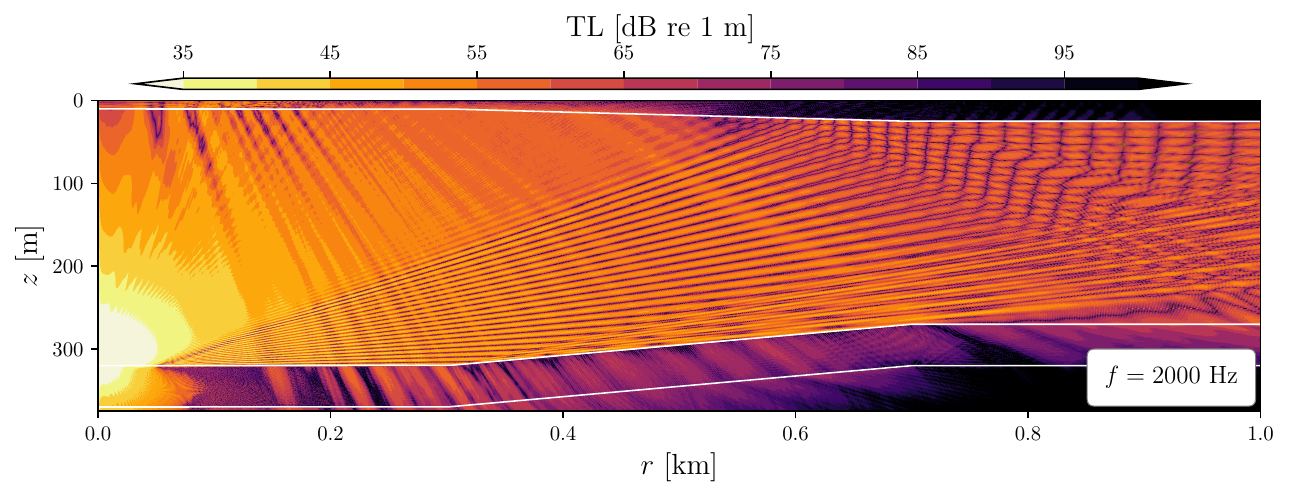}}
\caption{2D transmission loss (TL) plot for Example D at frequencies (a) 250 Hz and (b) 2000 Hz, computed with the PE using the square root operator approximated by the AAA algorithm. 
}\label{fig:exDclr}
\end{figure}

\begin{figure}
\centering
\includegraphics[width=\textwidth]{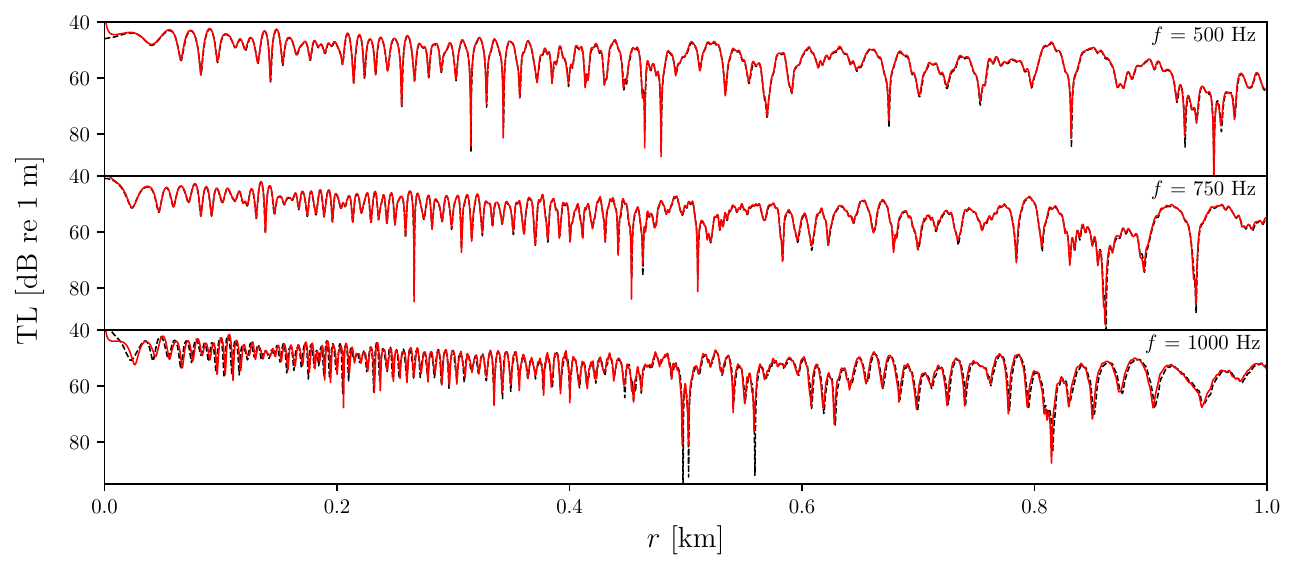}
\caption{Transmission loss (TL) at depth $z = 100$ m for Example D at (upper panel) $f=500$ Hz, (middle panel) $f=750$ Hz, and (lower panel) $f=1000$ Hz.
The solid red curves are results from simulations using the AAA algorithm to approximate the square root. 
The benchmark solution computed using the finite-element method is displayed in each panel as a black dashed line.
}\label{fig:exDTL}
\end{figure}

\begin{figure}
\centering
\includegraphics[width=.7\textwidth]{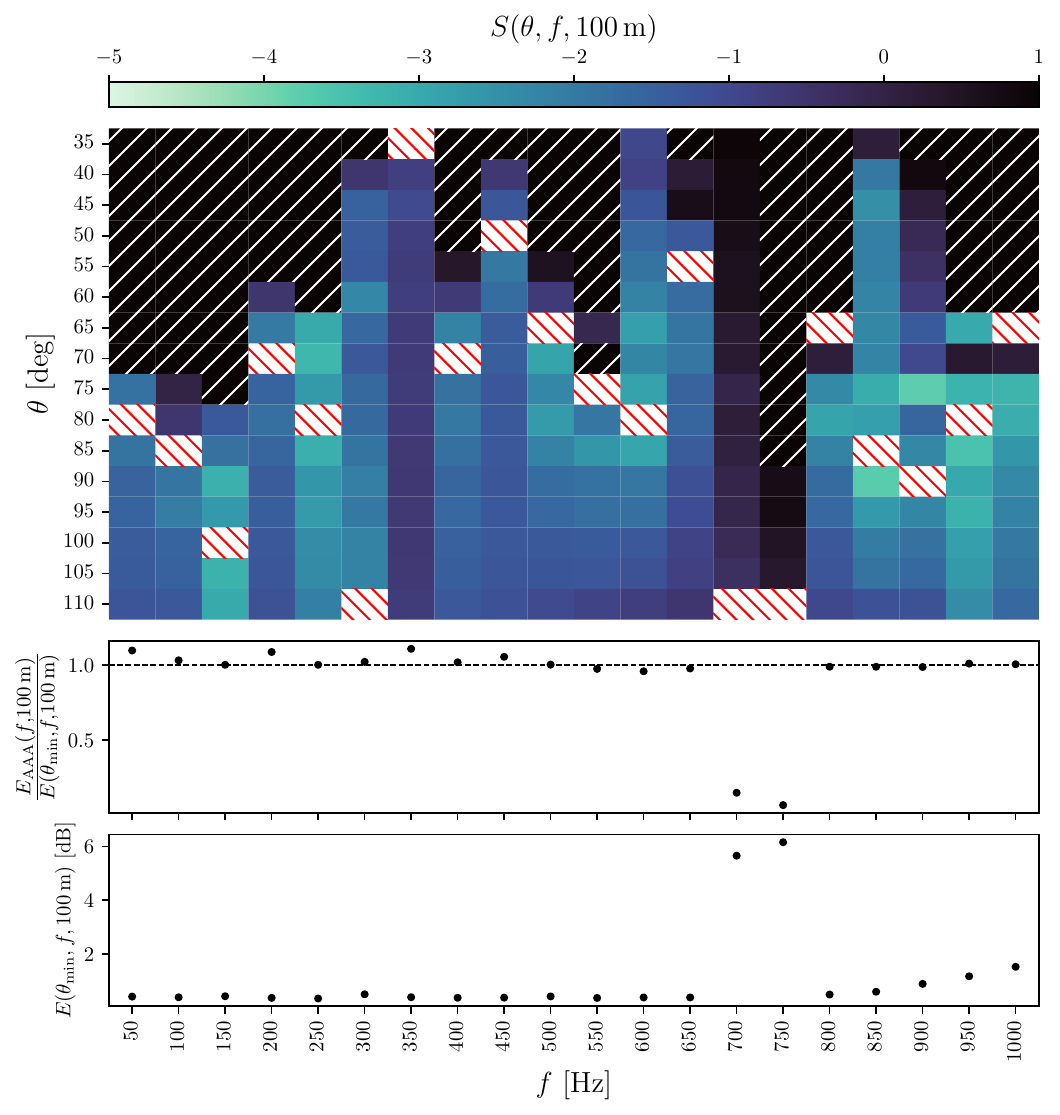}
\caption{Errors in transmission loss at $z=100$ m for Example D using approximations computed using 
(upper panel) the ROP method as a function of rotation angle $\theta$ and frequency $f$ relative to the optimal ROP $\theta$ value, and 
(middle panel) the AAA algorithm relative to the optimal ROP $\theta$ value. 
The lower panel shows the average error in transmission loss in dB using the ROP approximation with the optimal $\theta$ value.
The black squares with white hatches indicate combinations of $\theta$ and $f$ for which the simulation results diverged, while
the white squares with red hatches indicate the optimal value of $\theta$ for each frequency.
See text for more detail.}
\label{fig:thetaswpD}
\end{figure}

The environment for Example D is shown in Fig. \ref{fig:exDwg}.
The waveguide has an elastic layer of 10 m thickness for $r<300$ m and 25 m thickness at $r > 700$ m, with linearly increasing thickness from 300 m $<r<$ 700 m,
with density $\rho = 0.9$ g/cm$^3$, compressional wave speed $c_p=3500$ m/s, 
shear wave speed $c_s = 1750$ m/s, compressional attenuation $\alpha_p$ = 0.1 dB/$\lambda$, and shear attenuation $\alpha_s$ = 0.2 dB/$\lambda$, with $\lambda$ the wavelength.
This layer overlays a fluid medium ending at depth 320 m at $r<300$ m, 270 m at $r>700$ m, and linearly decreasing in depth from 300 m $<r<$ 700 m.
The fluid layer has density $\rho = 1.0$ g/cm$^3$ and sound speed $c_p = 1500$ m/s.
The next layer is an elastic medium of uniform thickness 50 m (the interface follows the bottom of the fluid medium), 
with density $\rho = 1.2$ g/cm$^3$, compressional wave speed $c_p=1700$ m/s, 
shear wave speed $c_s = 800$ m/s, compressional attenuation $\alpha_p$ = 0.1 dB/$\lambda$, and shear attenuation $\alpha_s$ = 0.2 dB/$\lambda$.
Finally, the rest of the computational domain is filled with an elastic halfspace with density $\rho = 1.5$ g/cm$^3$, compressional wave speed $c_p=2400$ m/s, 
shear wave speed $c_s = 1200$ m/s, compressional attenuation $\alpha_p$ = 0.2 dB/$\lambda$, and shear attenuation $\alpha_s$ = 0.4 dB/$\lambda$.
A source is placed at $z = 300$ m and the field is propagated out to 1 km range.

Fig. \ref{fig:exDclr} shows the 2D TL in this waveguide for a source of frequencies (a) 250 Hz and (b) 2000 Hz 
using the square root operator approximated with the AAA algorithm to propagate the field.
The simulations remain stable well into the kHz regime. 
Fig. \ref{fig:exDTL} shows TL curves at $z=100$ m for frequencies 500, 750, and 1000 Hz for Example D
for simulations using the AAA algorithm to approximate the square root.

Fig. \ref{fig:thetaswpD} shows the results of the sweep over values of $\theta$ from $35^\circ$ to $110^\circ$ in increments of 5 degrees, 
and frequencies 50 - 1000 kHz in 50 Hz increments.
Shown in the upper panel is the quantity $S(\theta,f,z)$ computed for $z=100$ m depth. 
The optimal rotation angle for each frequency for Example D does not coincide with that of Example C; 
the optimal $\theta$ to minimize overall error in the transmission loss is therefore waveguide dependent as well as being dependent on frequency.

The lower panel of Fig. \ref{fig:thetaswpD} shows the error of simulations using the AAA algorithm to approximate the square root normalized to the
error using the ROP with $\theta_\mathrm{min}$ at each frequency.
The lower panel shows the average error in transmission loss using the ROP method to approximate the square root with $\theta_\mathrm{min}$ at each frequency.
There are minor phasing errors that appear as frequency is increased, as can be seen in the 1000 Hz panel of Fig. \ref{fig:exDTL}, which causes the overall rise in average error with frequency for all simulations.

The approximated operators found using the AAA algorithm perform comparably to the most accurate simulations using the ROP method 
from the $\theta$ values sampled for Example D.
Notably, at 700 and 750 Hz, the ROP method does not give a convergent result at any rotation angle for the grid spacings used ($\Delta r = \lambda_0/16,\,\Delta z = \lambda_0/64$),
with a minimal average error of $\sim$6 dB/m,
while the simulations using the AAA approximation give an accurate result at both frequencies (c.f. the middle panel of Fig. \ref{fig:exDTL}).

The optimal value of $\theta$ shows no discernible pattern, and therefore it is not feasible to obtain the most accurate
result without performing a parameter sweep over the rotation angle and comparing to a benchmark solution.
In practice, PE simulations are carried out for environments in which it is not tractable to compute a solution using other methods (finite element method, etc.),
and therefore there is no method to determine the best $\theta$ for those waveguides.

\section{Summary and conclusions}

In numerical simulations of wave propagation using parabolic equation methods, depth operators of various functional forms need to be approximated.
Historically, Pad\'e approximations of operators with rotated branch cuts (referred to in this paper as the ROP method)
 have performed excellently in simulations of wave propagation in fluid media,
but have had issues handing mid- to high-frequency wave propagation in some waveguides with coupled fluid-elastic media.
In this paper, we have demonstrated that rational approximations of PE operators using the adaptive Antoulas-Anderson (AAA) algorithm allow for accurate and stable
simulation of wave propagation in waveguides where other approximation methods fail.

The AAA algorithm also allows for PE simulations of wave propagation in fluid-elastic waveguides 
to take advantage of the split-step Pad\'e approach, where the field is marched out in range using
the exponential form of the PE operator.
The approximation of the exponential operator incorporates range numerics into the coefficients of the rational function, 
meaning that the range step of the simulation can be multiple wavelengths while maintaining accuracy.
While this approach has been the standard in PE simulations of wave propagation in fluid waveguides for decades, 
the stability of the exponential operator was problematic for propagation in certain fluid-elastic waveguides.

 Accurate results using the ROP method were found to be highly dependent on a good choice of the rotation angle, 
 appropriately balancing the ability of the approximated operators to properly handle the evanescent
 spectrum while minimizing instabilities caused by introducing an exponentially growing component to propagating modes. 
 The AAA algorithm yields approximations that, without parameters, perform comparably to the ``best'' rotation angles, 
 and give convergent results where the ROP method may fail.

\section*{Acknowledgements}

This work was supported by the Office of Naval Research (ONR).
The authors thank L. T. Fialkowski and J. V. Lambers for helpful discussion and comments on the manuscript.

\appendix

\section{Pad\'e approximants}\label{apx:ratapx}

In this section, we overview the procedure for computing rational approximations of functions.
For a polynomial $T(x)$ of order $m+n$, there exists a rational function $R(x) = A(x)/B(x)$, with $A$ a polynomial
of order $m$ and $B$ a polynomial of order $n$, that agrees with $T(x)$ to the highest-possible order.
In particular, this means that the derivatives at each order match at a specific point (taken here to be 0):
\eq{
T^{(j)}(0) = R^{(j)}(0),\,\,\, j = 0,...,m+n\,.\nn
}
In order to find the Pad\'e approximant
of any function, we find its Taylor series at $x=0$ (Maclaurin series) of order $m+n$ and match $m+n$ derivatives with the rational
function.

For simplicity, take A and B each of order $n$. Define $N=2n$. Then $T(x) = \sum_{i=0}^Nt_ix^i$. $A$ and $B$ can be written similarly, with $a_i=b_i=0, i>n$. We begin with
\eq{
A(x) = T(x)B(x)\,,\nn
}
with $b_0 = 1$, which yields a set of equations for the coefficients of the polynomials $A$ and $B$ for matching orders of $x$:
\al{
a_0 &= t_0\nn \\
a_1-t_0b_1 &= t_1\nn \\
a_2 - t_1b_1-t_0b_2 &= t_2\nn \\
&\vdots\nn \\
a_N-t_{N-1}b_1- ... -t_0b_N &= t_N \nn\,,
}
summarized as 
\eq{
t_j = a_j - \sum_{i=1}^jt_{j-i}b_i\,,\,\,\,j=0,...,N.
}

This system of equations can be put into matrix form,

\eq{
\left[
\begin{array}{ccccccccccccccc}
\multicolumn{15}{c}{\multirow{6}[1]{*}{\Huge $C$}}\\
&&&&&&&&&&&&&&\\
&&&&&&&&&&&&&&\\
&&&&&&&&&&&&&&\\
&&&&&&&&&&&&&&\\
&&&&&&&&&&&&&&\\
\end{array}\right]
\left[
\begin{array}{c}
a_1\\
\vdots\\
a_n\\
b_1\\
\vdots\\
b_n
\end{array}\right]
=
\left[
\begin{array}{c}
t_1\\
t_2\\
\vdots\\
\vdots\\
t_{N-1}\\
t_N
\end{array}\right]\,,\nn
}
where the matrix $C$ represents the coefficients of the $a_j,\,b_j$.

When finding these Taylor-series coefficients
to large order, division of the derivatives by large factorials loses numerical precision. Therefore, a slightly different set of 
equations is solved, using the derivatives of the function rather than Taylor series coefficients. Substuting $f_j = j! t_j$,
\al{
f_j &= j!\left(a_j-\sum_{i=1}^j\frac{f_{j-i}}{(j-i)!}b_i\right)\\
&=j!a_j - \sum_{i=1}^j\frac{j!}{(j-i)!} f_{j-i}b_i \\ 
&=j!a_j - \sum_{i=1}^j i! \binom{j}{i} f_{j-i}b_i\,, \label{eq:pade_der}
}
for $j=0,...,N$. This set of equations is then put into a matrix equation, which can be solved to yield the polynomial coefficients.

To get these polynomial coefficients into the form of Eq. (\ref{eq:pade}),
we need to find the roots of $A(x)$ and $B(x)$. This can be done in many ways; for example, by utilizing Laguerre's root-finding algorithm. In this work, we find the roots 
from the companion matrix of the polynomials.
For a polynomial of order $n,\,g(x)=\sum_{i=0}^n a_ix^i$, its companion matrix is
\eq{
\begin{bmatrix}
    0   &   0       & \cdots         & 0    & -a_0/a_n\\
    1 &     0       & \cdots        & 0    & -a_1/a_n\\
    0 &     1       & \cdots         & 0    & -a_2/a_n\\
\vdots  & \vdots    & \ddots    & \vdots    & \vdots\\
    0 &     0       & \cdots    & 1         & -a_{n-1}/a_n
  \end{bmatrix} \nn
}
The roots of the polynomial $g$, denoted $r_i,\,i=1,...,n$, are the eigenvalues of the companion matrix. Then the polynomial can  be
expressed as
\eq{
    g(x) = c \prod_{i=1}^n(1+\gamma_i x)\,,
} 
with $\gamma_i = -1/r_i$, and $c = a_n \prod_{i=1}^n (-r_i) \equiv g(0)$.

To convert between the sum and product forms of the Pad\'e approximant,
\eq{
\prod_{j=1}^n\frac{1+\gamma_{j} q}{1+\mu_{j}q} \equiv 1 + \sum_{j=1}^n \frac{\alpha_{j} q}{1+\mu_{j}q}\,,\nn
}
one must carry out a similar procedure of matching coefficients of powers of $q$. Rearranging the above equation,
\eq{
\prod_{j=1}^n(1+\gamma_{j}q) = \prod_{j=1}^n(1+\mu_{j}q) + \sum_{j=1}^n \alpha_{j}q \prod_{k=1, k\neq i}^n (1+\mu_{k}q)\,.
}
This yields a system of equations 
which can be succinctly summarized as,
\al{
\sum_{j=1}^n \alpha_j &= \sum_{j=1}^n \gamma_j - \sum_{j=1}^n\mu_j \\
\sum_{j=1}^n\alpha_j\sum \binom{\{\mu\}_{\neq j}}{k - 1} &= \sum \binom{\{\gamma\}}{k} - \sum \binom{\{\mu\}}{k}\,,\label{eq:prodtosum}
}
with $k=2...n$, where $\{\gamma\}$ and $\{\mu\}$ are the sets of coefficients, $\{\mu\}_{\neq j}$ is the set of $\mu$ not including $\mu_j$, and $\sum \binom{\{S\}}{k}$ 
indicates a sum of all combinations of $k$ elements of the set $\{S\}$.

\section{Standard approaches for stable Pad\'e approximations}\label{apx:hist}

In this section, we briefly overview typically used methods of stabilizing rational approximations of the PE operators.
To briefly explain the need for stabilizing rational approximations,
we will begin by looking at the case of the square root operator.
The rational approximation of this operator has  closed forms for its Pad\'e coefficients,
\eq{
\sqrt{1+q} \approx 1 +  \sum_{i=1}^n \frac{\alpha_{j}q}{1+\mu_{j}q} = \prod_{i=1}^n \frac{1+\gamma_{j} q}{1+\mu_{j}q} \,,\nn
}
with
\al{
\alpha_{j}&=\left(\frac{2}{2n+1}\right)\sin^2\left(\frac{j\pi}{2n+1}\right)\,,\\ 
\mu_{j}&=\cos^2\left(\frac{j\pi}{2n+1}\right)\,, \\ 
\gamma_{j}&=\sin^2\left(\frac{j\pi}{2n+1}\right)\,.
}

While this approximation gives excellent agreement with the square root function for $q\geq-1$, 
it is poorly behaved for $q < -1$; real-valued coefficients are not able to reproduce the correct behavior of the function
when it has an imaginary component. In wave propagation, the region $q <-1$
represents the evanescent spectrum, and an accurate approximation is essential
to correctly modeling propagation.
 Evanescent modes that are treated as propagating modes, rather than annihilated, will build and result in unstable and inaccurate simulations.

In addition to the approximated function not having an imaginary component,
the problem of accurately modeling this region is compounded by the fact that the branch cut of the square-root operator lies on the real line, on which
we are evaluating the function. In terms of the Pad\'e approximant, there are poles along the negative real axis at points $q = -1/\mu_{j}$. 
 The Pad\'e approximation using the coefficients above, therefore, is not sufficient for application in PE propagation methods.

There are two typical approaches for fixing these issues. 

\subsection{Constraint equations}
The first is to replace one (or more -- though one is sufficient for acoustic problems) of the Pad\'e approximant derivative equations with a constraint
equation that moves the poles into the complex plane (i.e. gives a complex component to the $\mu_{j}$ and also move the branch cut from the negative real line)\cite{collins1991.1}. 
Concretely, one possibility for one constraint is to replace the $N$th equation of Eq. (\ref{eq:pade_der}),
\eq{
f_N(q_0) = g(q_0) - (g_0(q_0) + i \epsilon)\,, \label{eq:stab_constraint1}
}
where $g$ is the perturbed Pad\'e approximant of the function, $g_0$ is the unperturbed approximant, $\epsilon$ is the perturbation into the complex plane,
and $q_0$ lies on the negative real axis.

In practice, this amounts to solving for the system twice. First, we solve system of equations specified by Eq. (\ref{eq:pade_der}), which yields the polynomial coefficients
$\{a^{(0)}\}$, $\{b^{(0)}\}$ for the rational approximation
\eq{
g_0(q)  = \frac{\sum_{i=0}^n a^{(0)}_iq^i}{\sum_{i=0}^n b^{(0)}_iq^i}\,. \nn
}
Second, the system of equations is set up once more
with the last equation of the system replaced by Eq. (\ref{eq:stab_constraint1}), with explicit form 
\eq{
\sum_i^na_iq_0^i - (g_0(q_0) + i \epsilon) \sum_i^nb_iq_0^i = 0 \,, \nn
}
where we use the previous solution to evaluate $g_0(q_0)$.
This second solve yields the perturbed coefficients $\{a\}$ and $\{b\}$.
This method is applicable only to the $(1+cq)^\nu$ form of the operator.

Another possible constraint equation is to replace the $N$th equation of Eq.~(\ref{eq:pade_der}) with a fixed value at point $q_0$\cite{collins1993}. 
This fixed value could either be the function evaluated at that point, or a value specifically chosen
for stability, i.e. $g(q_0) = f(q_0)$ or $g(q_0) = 0$. 
The latter is typically used for the exponential and self-starter operators.
This requires only a single solve of the system.  

\subsection{Rotated operators}
The second approach, as detailed in Milinazzo, et al.\cite{milinazzo1997}, is to rotate the 
square root operator such that its branch cut no longer lies on the negative real line.
Define the rotated coordinate $\tilde q = e^{-i\theta}(1 + q)-1$, with $\theta$ the rotation angle. 
Then,
\eq{
\sqrt{1+q} = e^{i\theta/2}\sqrt{1+\tilde q} \approx e^{i\theta/2}\left(1+\sum_{j=1}^n\frac{\alpha_{j}\tilde q}{1 + \mu_{j} \tilde q}\right)\,, \nn
}
where the coefficients $\{\alpha\},\{\mu\}$ are those presented earlier for the unrotated operator.

Substituting for $\tilde q$ gives
\al{
f(q) &= \sqrt{1+q} \approx e^{i\theta/2}\left(1 + \sum_{j=1}^n  \frac{\alpha_{j}(e^{-i\theta}(q+1) - 1)}{1 + \mu_{j}(e^{-i\theta}(q+1) - 1)}\right) \nn \\
& = e^{i\theta/2}\left(1+\sum_{j=1}^n \frac{\alpha_{j}(e^{-i\theta} - 1)}{1 + \mu_{j}(e^{-i\theta}-1)}\right) + \sum_{j=1}^n\frac{\tilde\alpha_{j}q}{1+\tilde\mu_{j}q}\,, \nn
}
with
\eq{
\tilde\alpha_{j} = \frac{e^{-i\theta/2}\alpha_{j}}{(1+\mu_{j}(e^{-i\theta} - 1))^2}\,,
~~~~\tilde\mu_{j} = \frac{e^{-i\theta}\mu_{j}}{1+\mu_{j}(e^{-i\theta} - 1)}\,\nn.
}
The first term is the approximation of
\eq{
e^{i\theta/2}f(e^{-i\theta} - 1) = e^{i\theta/2} e^{-i\theta/2} = 1 \nn\,,
}
so
\eq{
f(q) = \sqrt{1+q} \approx 1 + \sum_{i=1}^n\frac{\tilde\alpha_{j}q}{1+\tilde\mu_{j}q}\,.
}
These coefficients are particularly convenient to use in simulations as they have closed forms;
the coefficients for approximating the rotated operator are a simple transformation of those for the unrotated operator.

We now carry out this procedure for the generalized PE operator in Eq. (\ref{eq:peop}).
As above, define $\tilde q = e^{-i\theta}(1 + q)-1$. Then the rotated operator $\tilde f$ is
\al{
\tilde f(\tilde q,\theta) &\equiv f(e^{i\theta}(1 + \tilde q)-1) \\
&=\exp\{i\sigma(-1 + e^{i\theta/2}\sqrt{1+\tilde q}) + \delta\ln[e^{i\theta}(1+\tilde q)] + \nu \ln[1-c+ce^{i\theta}(1+\tilde q)]\} \,. 
}
We first find the Pad\'e approximation of $\tilde f$ at the point $\tilde q = 0$ for some fixed value of $\theta$,
\eq{
\tilde f(\tilde q, \theta) \approx \tilde f(0, \theta) \left(1+\sum_{j=1}^n\frac{\tilde \alpha_{j}\tilde q}{1 + \tilde \mu_{j} \tilde q}\right) = \tilde f(0, \theta) \prod_{j=1}^n \frac{1+\tilde \gamma_{j} \tilde q}{1+\tilde \mu_{j} \tilde q} \nn \,,
} 
where the tildes on the coefficients indicate that they are for the {\em rotated} operator.
These coefficients can be found using the method detailed in Apx. \ref{apx:ratapx}.

Putting this operator back in terms of $q$,
\al{
f(q) &\approx \tilde f(0, \theta) \prod_{j=1}^n \frac{1+\tilde \gamma_{j}  (e^{-i\theta}(1 + q)-1) }{1+\tilde \mu_{j} (e^{-i\theta}(1 + q)-1)} \nn\\
&= \tilde f(0, \theta) \left(\prod_{j=1}^n \frac{1+\tilde \gamma_{j} (e^{-i\theta}-1)}{1+\tilde \mu_{j} (e^{-i\theta}-1)}\right)\left( \prod_{j=1}^n \frac{1+\bar \gamma_{j} q}{1+\bar \mu_{j}q}\right) \\
& \equiv P \left( \prod_{j=1}^n \frac{1+\bar \gamma_{j} q}{1+\bar \mu_{j}q}\right)\nn\,,
}
with
\eq{
\bar \gamma_{j} = \frac{e^{-i\theta}\tilde\gamma_{j}}{1+\tilde\gamma_{j}(e^{-i\theta} - 1)}\,,
 ~~~~\bar\mu_{j} = \frac{e^{-i\theta}\tilde\mu_{j}}{1+\tilde\mu_{j}(e^{-i\theta} - 1)}\,\nn.
}
The pre-factor $P$ is the approximation of $\tilde f(e^{-i\theta}-1,\theta) \equiv f(0) = 1$, so 
\eq{
f(q) \approx \prod_{j=1}^n \frac{1+\bar \gamma_{j} q}{1+\bar\mu_{j}q}\,.
}

When $\delta=1/2,\,\sigma=0,\,\nu=0$, this gives back the result for the rotated square root operator.
The generalized operator does not have a closed form for its coefficients, and they must be numerically computed
for each set of parameters.

\section{Derivatives of rotated operator}\label{apx:deriv}

For ease of numerical implementation of the method in Apx. \ref{apx:hist}, we detail a recursive method for calculation of the derivatives of the operator $\tilde f$ here.

First, we take the natural logarithm of the operator,
\eq{
\ln(\tilde f) = i\sigma(-1 + e^{i\theta/2}\sqrt{1+\tilde q}) + \delta\ln[e^{i\theta}(1+\tilde q)] + \nu \ln[1-c+ce^{i\theta}(1+\tilde q)]\,.
}
The derivative of this function with respect to $\tilde q$ is,
\eq{
\pdif{\ln(\tilde f)}{\tilde q} = \frac{\tilde f'}{\tilde f}\equiv d(\tilde q) = \frac{i \sigma e^{i\theta/2}}{2(1+\tilde q)^{1/2}} + \frac{\delta}{1+\tilde q} + \frac{\nu c e^{i \theta}}{1-c+ce^{i\theta}(1+\tilde q)} \,. \nn
}
Then,
\eq{
\tilde f_j = \sum_{k=0}^{j-1} \binom{j-1}{k} d_{(j-1-k)}\tilde f_{k}\,.
}
The derivatives of function $d$ are
\al{
d_{j} =  \left(\prod_{k=0}^j \left(\frac{1}{2} - k \right) \right) \frac{i\sigma e^{i\theta/2}}{(1+\tilde q)^{1/2+j}} + (-1)^j j! \frac{\delta}{(1+\tilde q)^{j+1}} + (-1)^j j!  \frac{\nu c^{j+1} e^{(j+1)i\theta}}{\left(1-c+ce^{i\theta}(1+\tilde q)\right)^{j+1}}\,.
}

\bibliographystyle{ws-jtca}
\bibliography{references}

\end{document}